\documentclass[aps,prx,numerical,reprint,amsmath,amssymb,superscriptaddress,longbibliography,nofootinbib,onecolumn]{revtex4-2}
\usepackage{mathtools, amsthm, mathrsfs}
\usepackage[braket, qm]{qcircuit}
\usepackage{siunitx, physics}
\usepackage{float, placeins}
\usepackage{bbm, bm}
\usepackage{xparse}
\usepackage{lipsum, comment} \usepackage{graphicx, svg} \graphicspath{{Figures/}}
\usepackage{soul}
\usepackage{times}
\usepackage{comment}
\usepackage[normalem]{ulem}
\usepackage{algorithm2e}
\usepackage{bibunits}
\defaultbibliographystyle{apsrev4-2}
\defaultbibliography{bibfile}
\usepackage{titletoc, tocloft} \usepackage{dcolumn, tabularx, makecell} 
\usepackage{xcolor}
\definecolor{linkblue}{rgb}{0.0,0.0,0.6}
\usepackage[colorlinks=true,
            linkcolor=linkblue,
            citecolor=linkblue,
            urlcolor=linkblue]{hyperref}

\newcommand{\tc}[1]{{\textcolor{orange}{[TC: #1]}}}

\renewcommand{\vec}[1]{\bm{#1}}
\newcommand{\mat}[1]{\mathbf{#1}}
\newcommand{\scale}{1}

\newcommand{\Kr}{{K_\mathrm{r}}}
\newcommand{\Lr}{{L_\mathrm{r}}}

\newcommand{\suppnoteref}[1]{Supplemental Material Section\,\ref{#1}}

\DeclareMathOperator{\mean}{\mathbb{E}}
\DeclareMathOperator{\snr}{\mathrm{SNR}}

\DeclareMathOperator{\Var}{\mathrm{Var}}
\DeclareMathOperator{\Cov}{\mathrm{Cov}}

\newcommand{\eqnref}[1]{Eq.~\eqref{#1}}
\newcommand{\Eqnref}[1]{Equation~\eqref{#1}}
\newcommand{\suppEqref}[1]{Eq.~\eqref{#1}}
\newcommand{\SuppEqref}[1]{Equation~\eqref{#1}}

\newcommand{\figref}[1]{Fig.~\ref{#1}}

\NewDocumentCommand{\figpanelref}{s m m g}{\IfNoValueTF{#4}
    {Fig.~\ref{#2}(#3)}
    {\IfBooleanTF{#1}
        {Figs.~\ref{#2}(#3)--\ref{#2}(#4)}
        {Figs.~\ref{#2}(#3) and~\ref{#2}(#4)}}}

\NewDocumentCommand{\Figpanelref}{s m m g}{\IfNoValueTF{#4}
    {Figure~\ref{#2}(#3)}
    {\IfBooleanTF{#1}
        {Figures~\ref{#2}(#3)--\ref{#2}(#4)}
        {Figures~\ref{#2}(#3) and~\ref{#2}(#4)}}}

\newcommand{\suppfigref}[1]{Fig.~\ref{#1}}
\newcommand{\Suppfigref}[1]{Figure~\ref{#1}}

\NewDocumentCommand{\suppfigpanelref}{s m m g}{\IfNoValueTF{#4}
    {Fig.~\ref{#2}(#3)}
    {\IfBooleanTF{#1}
        {Figs.~\ref{#2}(#3)--\ref{#2}(#4)}
        {Figs.~\ref{#2}(#3) and~\ref{#2}(#4)}}}

\NewDocumentCommand{\Suppfigpanelref}{s m m g}{\IfNoValueTF{#4}
    {Figure~\ref{#2}(#3)}
    {\IfBooleanTF{#1}
        {Figures~\ref{#2}(#3)--\ref{#2}(#4)}
        {Figures~\ref{#2}(#3) and~\ref{#2}(#4)}}}

\newcommand{\figtitle}[1]{{#1}}
\NewDocumentCommand{\figpanel}{s m g}{\IfBooleanTF{#1}
    {\IfNoValueTF{#3}{(#2)}{(#2)--(#3)}}{\IfNoValueTF{#3}{(#2)}{(#2),~(#3)}}}

\begin{document}

\title{Measurement-Adapted Eigentask Representations for Photon-Limited Optical Readout}

\author{Tianyang~Chen}
\thanks{These authors contributed equally to this work.}
\affiliation{Department of Electrical and Computer Engineering, Princeton University, Princeton, NJ 08544, USA}

\author{Mandar~M.~Sohoni}
\thanks{These authors contributed equally to this work.}
\affiliation{School of Applied and Engineering Physics, Cornell University, Ithaca, NY 14853, USA}

\author{Saeed~A.~Khan}
\affiliation{School of Applied and Engineering Physics, Cornell University, Ithaca, NY 14853, USA}

\author{J\'{e}r\'{e}mie Laydevant}
\affiliation{School of Applied and Engineering Physics, Cornell University, Ithaca, NY 14853, USA}
\affiliation{USRA Research Institute for Advanced Computer Science, Mountain View, CA 94035, USA}

\author{Shi-Yuan~Ma}
\affiliation{School of Applied and Engineering Physics, Cornell University, Ithaca, NY 14853, USA}

\author{Tianyu~Wang}
\affiliation{School of Applied and Engineering Physics, Cornell University, Ithaca, NY 14853, USA}

\author{Peter~L.~McMahon}
\email{pmcmahon@cornell.edu}
\affiliation{School of Applied and Engineering Physics, Cornell University, Ithaca, NY 14853, USA}
\affiliation{Kavli Institute at Cornell for Nanoscale Science, Cornell University, Ithaca, NY 14853, USA}

\author{Hakan~E.~T\"ureci}
\email{tureci@princeton.edu}
\affiliation{Department of Electrical and Computer Engineering, Princeton University, Princeton, NJ 08544, USA}

\newcommand{\docdate}{May 10, 2026}
\date{\docdate}

\begin{bibunit}[apsrev4-2]
\nocite{apsrev42Control}
\begin{abstract}
Optical readout in low-light imaging is fundamentally limited by measurement noise, including photon shot noise, detector noise, and quantization error. In this regime, downstream inference depends not only on the optical front end, but also on how noisy high-dimensional sensor measurements are represented before classification or decision-making. Here we show that eigentasks provide a measurement-adapted representation for optical sensor outputs by ordering readout features according to their resolvability under noise. Using experimental data from a lens-based optical imaging system and a reanalysis of published data from a single-photon-detection neural network, we find that eigentask representations frequently outperform standard baselines including principal component analysis and filtering-based compression. The advantage is most pronounced in photon-limited, few-shot, and higher-difficulty classification regimes. In few-shot MPEG-7 classification, for example, the advantage over other methods reaches about 10 percentage points as the number of classes increases. In these settings, eigentasks yield more informative low-dimensional features and improve sample-efficient downstream learning. These results identify measurement-adapted representation as a promising strategy for optical inference when photon budget, acquisition time, and task complexity are constrained.
\end{abstract}

\maketitle

\section{Introduction}

Optical sensors form the interface between physical scenes and digital inference. In a range of important applications, however, the limiting resource is not the number of detector channels or the complexity of the optical front end, but the photon budget available per measurement. This regime arises in settings such as live-cell imaging, where illumination must be limited to avoid photobleaching and phototoxicity~\cite{ettinger2014fluorescence, laissue2017assessing, icha2017phototoxicity, fazel2024fluorescence}, and in high-speed optical measurements where exposure time is constrained~\cite{klein2017highspeeda, zhang2024pixelwise, platisa2023highspeed}. Under these conditions, each detector element collects only a small number of photons, and the resulting readout becomes strongly stochastic, with signal-to-noise ratios (SNRs) set by photon shot noise together with detector and quantization noise~\cite{healey1994radiometric, elgamal2005cmos, janesick1987scientific, bigas2006review, foi2008practical}. As a result, under a fixed or constrained photon budget, increasing the number of pixels or measurement channels does not automatically translate into better inference~\cite{farrell2006resolution, chan2024resolution}. The challenge shifts from acquiring more data to extracting reliable information from noisy, high-dimensional optical measurements~\cite{schott2016impact, dodge2016understanding, hendrycks2018benchmarking, rapp2020advances, bian2023highresolution}.

A central challenge in this setting is how to convert noisy sensor outputs into a lower-dimensional representation that preserves task-relevant information. Standard approaches, including spatial filtering, Fourier-domain truncation, and principal component analysis (PCA), are widely used for denoising or compression~\cite{gonzalez2009digital, cunningham2015linear, rasti2018noise}, but they are generally not explicitly constructed to reflect the spatio-temporal measurement statistics of a specific optical readout process. Consequently, the components they retain are not necessarily those that remain most observable under the actual noise structure of the sensor, particularly in photon-limited regimes.

Eigentasks offer an alternative. Introduced in the context of qubit-based quantum systems as computational substrates~\cite{hu2023tackling} (see also Ref.~\cite{polloreno2025restrictions, wang2025advancing}), eigentasks define a basis in the space of a complete set of projective measurement operators to maximize the signal-to-noise ratio of the resulting feature components with respect to inputs drawn from a specified prior. Conceptually, eigentasks represent a native set of functions of the input that a given quantum dynamical system can approximate with minimal error when projectively measured.

Here we show that the same principle can be used to construct a measurement-adapted representation of optical sensor readout for downstream inference under photon-limited conditions. In their original formulation, eigentasks provide notable advantage for projective measurements subject to multiplicative noise, and it is therefore not a priori clear that these benefits carry over to optical readout. We demonstrate, through application to distinct optical sensing architectures, that eigentasks define an effective representation layer for optical measurements by ordering features according to their observability under noise. This perspective extends the role of eigentasks from tools for analyzing physical computation to a general strategy for extracting information from noisy optical measurements, suggesting a principled approach to the design of optical inference pipelines.

\begin{figure}[!htbp]
    \centering
    \includegraphics[scale=\scale]{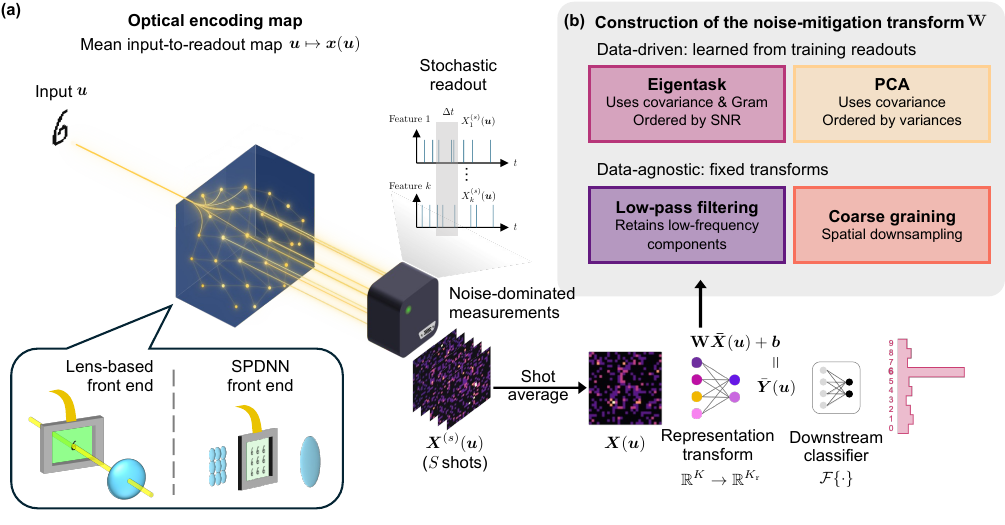}
    \caption{
    \figtitle{Image classification through eigentask-based noise mitigation.}
    \figpanel{a} A schematic of the optical classification pipeline. Inputs are encoded in light, processed by an optical front end (lens-based or SPDNN, as considered in this work), and measured through stochastic readout noise. Repeated shots are collected and averaged to form the sensor readout used for subsequent processing, including a representation transform for noise mitigation and a downstream classifier $\mathcal{F}\{\cdot\}$ for label prediction. Here, $\vec{u}\mapsto\vec{x}(\vec{u})$ denotes the optical encoding map, which is a deterministic mean input-to-readout map implemented by the optical front end and the readout device in the infinite-shot limit (for details, see \suppnoteref{si:lens_transform} and \ref{si:spdnn_transform}).
    \figpanel{b} Comparison of how the transform $\mat{W}$ is constructed in the four noise-mitigation methods. Eigentask and PCA transforms are learned from training readouts: eigentasks are noise-aware and ordered by signal-to-noise ratio (SNR), whereas PCA is ordered by variance. By contrast, low-pass filtering and coarse graining are fixed, data-agnostic transforms.
    }
    \label{fig:schematic}
\end{figure}

In this work, we evaluate eigentask representations on two distinct optical readout settings: a lens-based imaging system measured with an electron-multiplying charge-coupled device (EMCCD) camera, and a single-photon-detection neural network (SPDNN) architecture. Using experimental data and validated simulations, we find that eigentasks frequently outperform standard baselines, including PCA and filtering-based compression, with the strongest gains in photon-limited and few-shot regimes. The advantage becomes more pronounced as classification difficulty increases, as evidenced in classification problems with a larger number of classes. These results suggest that, in photon-limited optical systems, performance can be improved not only by redesigning the optical front end, but also by redesigning how its output is represented prior to inference.

\section{Noise in measurement of optical systems}

For a physical sensor subject to an input $\vec{u}$, we define $\vec{X}(\vec{u})$ as its corresponding output in one experiment (``shot"); for a sensor with $K$ observable readout features, $\vec{X}$ will be a $K$-dimensional vector. Even for a fixed input $\vec{u}$, the sensor output will generally vary over repeated samples or measurements of the sensor, with the sample-to-sample variability determined by the strength of the noise in readout. Henceforth we will refer to the \emph{shot-averaged output} $\bar{\vec{X}}(\vec{u})$ over $S$ shots, described as a stochastic variable 
\begin{align}
    {\bar{\vec{X}}}(\vec{u}) = \vec{x}(\vec{u}) + \frac{1}{\sqrt{S}} \vec{\zeta}(\vec{u}),
    \label{eq:X}
\end{align} 
where $\mean_{\vec{\mathcal{X}}}[{\bar{\vec{X}}}(\vec{u})] = \vec{x}(\vec{u})$ describes the mean of sensor outputs, while $\vec{\zeta}(\vec{u})$ is a zero-mean stochastic vector that captures all its noise characteristics and $\mean_{\vec{\mathcal{X}}}[\cdot]$ denotes the expectation over sampling of outputs with $\vec{\mathcal{X}}(\vec{u})\equiv\{\vec{X}^{(s)}(\vec{u})\}_{s\in[S]}$ representing the measured feature vectors in $S$ shots. \Eqnref{eq:X} is extremely general, describing static stochastic outputs from any physical sensor. In a physical setting, individual static samples are inevitably acquired through time integration of a physical observable, potentially using a separately optimized filter, see Ref.~\cite{turin1960introduction, olcer2017adaptive, walter2017rapid, khan2024practical}. In this work we do not consider the optimization of the filter.

In the optical setting considered here, the input $\vec{u}$ is encoded into a propagating beam and processed by an optical front end. We denote by $\vec{u} \mapsto \vec{x}(\vec{u})$ the deterministic mean input-to-readout map implemented by the optical front end and the measurement device in the infinite-shot limit, and refer to it as the \emph{optical encoding map} (for details, see \suppnoteref{si:lens_transform} and \ref{si:spdnn_transform}, which include Ref.~\cite{goodman1969introduction}).
In the single-lens experiments, this map is fixed by phase modulation on the SLM, propagation through the single lens, and pixel-integrated intensity readout. In the SPDNN case, by contrast, the optical front end is a trained, task-specific layer of an optical neural network, with single-photon detectors serving as the readout. 
Both setups are illustrated in the inset of \figpanelref{fig:schematic}{a}. In either case, the experimentally observed readout fluctuates around the mean $\vec{x}(\vec{u})$ with noise $\vec{\zeta}(\vec{u})$ whose statistics depend on the readout device and the measurement protocol. For an optical sensor employing EMCCD cameras as readout, for example, $X_k$ represents the integrated light intensity (over a fixed, experimentally chosen integration window) on the $k$-th camera pixel, while $\zeta_k$ characterizes the noise arising from both Poissonian nature of incident photons and multiplication process within the camera (see \suppnoteref{si:lens_noise}, which includes Ref.~\cite{toninelli2020quantum, ProEM_HS_512BX3}). For single-photon detectors $X_k$ is binary-valued, representing the detection or non-detection of a photon, with $\zeta_k$ describing binomial noise (see \suppnoteref{si:spdnn_setup_noise}). In both cases, a fixed integration time is assumed per shot.

In standard imaging, the objective is typically to obtain a high-SNR estimate of the expected optical sensor output, or image, denoted $\vec{x}(\vec{u})$. In many modern applications, however, the quantity of interest is not the image itself but a specific property of the input $\vec{u}$~(see \figpanelref{fig:schematic}{a}). In the settings considered here, $\vec{u}$ may represent, for example, a vector of pixel values on a SLM. A particularly relevant application is machine-vision classification. Each input $\vec{u} \sim \mathcal{D}$ is drawn from a dataset subject to a prior $\mathcal{D}$ and associated with a class label $c \in \{1,\ldots,C\}$, where typically the number of inputs greatly exceeds the number of classes. Each class could for instance correspond to different objects that may be encountered by a self-driving car. Given an input $\vec{u}^{(c)}$, the task is to predict a label $c_{\mathrm{pred}}$ from the corresponding stochastic sensor outputs $\vec{X}(\vec{u}^{(c)})$, with correct classification defined by $c_{\mathrm{pred}} = c$. Formally, the objective of classification tasks is to compute the task-dependent (and generally nonlinear) function $\mathcal{F}$ that outputs the predicted label $\mathcal{F}\{\bar{\vec{X}}(\vec{u}^{(c)})\} = c_{\mathrm{pred}}$ when applied to shot-averaged sensor measurements, rather than to reconstruct the image itself.

As the sensor output is stochastic, the accuracy of the prediction will be constrained by the sampling noise $\vec{\zeta}$: for sufficiently large stochasticity, $c_{\mathrm{pred}}$ does not faithfully reveal the correct label $c$, thereby limiting the classification accuracy. For supervised machine-learning tasks, a further complication arises. The function $\mathcal{F}\{ \cdot \}$ that maps stochastic sensor data to a predicted class label is \textit{a priori} unknown, and must itself be learned from noisy sensor data obtained for known inputs (the process of \textit{training}). While some amount of noise in the training data can prove beneficial by serving as a natural regularizer~\cite{bishop1995training}, too much noise will lead to increased generalization errors, ultimately restricting the ability to learn.

Here, we ask the question: can the measurement noise $\vec{\zeta}$ of the optical sensor output be reduced to mitigate these effects? The SNR of optical measurements is fundamentally governed by two factors: (i) the incident optical power and (ii) the number of shots $S$, or equivalently the exposure time. In many applications, the available optical power is tightly constrained; for example, when probing photosensitive or biological samples, the incident power must be limited to avoid perturbing the system under observation~\cite{laissue2017assessing}. Likewise, in settings that require rapid readout for feedback and control, such as autonomous driving~\cite{rapp2020advances}, the number of shots and exposure time are restricted by latency requirements. As a result, the sensor often operates under a limited photon budget and hence in a low-SNR regime, making the efficient extraction of information from noisy measurements essential.

\subsection{Noise-mitigation techniques}
\label{sec:noisemit}

In this work, we investigate whether applying noise-mitigation transformations to optical sensor outputs can improve learning performance in machine-vision classification tasks. To this end, we consider a general \textit{linear} transformation $\mat{W}$ with bias $\vec{b}$ applied to the sensor output prior to both training, and inference on unseen data. Formally,
\begin{align}
    \bar{\vec{Y}}(\vec{u}^{(c)})=\mat{W}\bar{\vec{X}}(\vec{u}^{(c)}) + \vec{b},\quad  c_{\mathrm{pred}} = \mathcal{F}\{ \bar{\vec{Y}}(\vec{u}^{(c)}) \},
    \label{eq:noisemit}
\end{align}
where $\mat{W} : \mathbb{R}^K \to \mathbb{R}^{\Kr}$ and $\vec{b} \in \mathbb{R}^{\Kr}$ define an affine map from the $K$-dimensional, $S$-shot averaged features $\bar{\vec{X}}(\vec{u}^{(c)})$ to noise-mitigated features $\bar{\vec{Y}}(\vec{u}^{(c)}) \in \mathbb{R}^{\Kr}$ with $\Kr \leq K$. As discussed before, the map $\mathcal{F}\{\cdot\}$ is a downstream trainable transformation that assigns predicted class labels from the processed features, see \figpanelref{fig:schematic}{a}. Thus, $\Kr$ denotes the number of retained features after noise mitigation. The transformation $\mat{W}$ with bias $\vec{b}$ is designed to improve generalization performance relative to training and inference on raw data (which corresponds to $\mat{W} = \mathbb{I}$ and $\vec{b} = \vec{0}$). When $\Kr < K$, the transformation also performs dimensionality reduction, establishing a close connection between noise mitigation and feature compression. Importantly, we focus on a model-free approach that does not require knowledge of the underlying input-to-readout map $\vec{u}\mapsto\vec{x}(\vec{u})$ and the noise structure that generate the raw and averaged features $\bar{\vec{X}}(\vec{u})$, which may be unknown or only partially characterized in physical sensing systems. We will show that even without explicit knowledge of this transformation, a data-driven noise-mitigation procedure can lead to improved learning performance in machine-vision tasks.

A widely used class of noise-mitigation techniques is filtering~\cite{gonzalez2009digital, milanfar2025denoising, goyal2020image}.
For two-dimensional imaging data, in particular for readout with $K = L \times L$ pixels, a simple approach is spatial coarse graining, which produces an image of reduced dimension $\Kr = \Lr \times \Lr$ by replacing groups of neighboring pixels with their weighted average (see \figpanelref{fig:mnist_main}{a}). A related method is Fourier-domain low-pass filtering, which applies a Fourier transform to the image followed by a low-pass filter to suppress high-frequency components associated with noise (see \figpanelref{fig:mnist_main}{a}). Both approaches are model-free and largely agnostic to the underlying data, aside from a small number of tunable hyperparameters.

While this simplicity makes the aforementioned approaches straightforward to implement, it is natural to ask whether improved noise mitigation can be achieved by exploiting statistical structure in the measured data. A canonical example is principal component analysis (PCA), which has been widely used to identify informative low-dimensional structure in high-dimensional datasets for compression and representation learning~\cite{jolliffe2016principal, greenacre2022principal}. In machine-learning applications, PCA is commonly employed to reduce dimensionality while retaining directions of largest variance. Although PCA does not rely on an explicit forward model, it is data-dependent, being defined by the empirical covariance of the measured features. Formally, the principal components are the $K$ eigenvectors $\{\vec{v}^{(k)}\}$ of the empirically estimated covariance matrix $\mat{C}$ of the $K$-dimensional sensor data. These components are ordered according to their associated eigenvalues, $\sigma_1^2 \geq \sigma_2^2 \geq \cdots \geq \sigma_K^2$, which quantify the variance captured along each direction (see Appendix\,\ref{app:methods}). For noise mitigation, the transformation in \eqnref{eq:noisemit} is chosen as $\mat{W}_{\mathrm{PCA}}^{\mathsf{T}} = (\vec{v}^{(1)} \ \cdots \ \vec{v}^{(\Kr)})$, corresponding to projection of the measured data onto the subspace spanned by the leading $\Kr$ principal components (see \figpanelref{fig:mnist_main}{a}).

The eigentask framework, introduced in Ref.~\cite{hu2023tackling} in the context of physical computing, where computation is performed by the dynamics of a physical device in response to input-encoded stimuli, provides a systematic approach for extracting information-rich features from noise-dominated measurements. As in PCA, eigentask learning employs a linear transformation that projects measured data onto an ordered basis. However, whereas principal components are ranked by the variance they capture, eigentasks are ordered according to the signal-to-noise ratio (SNR) of their observation. Formally, the eigentasks $\{\vec{r}^{(k)}\}$ form a set of $K$ vectors for $K$-dimensional data, ordered by $\alpha_k^2$ such that $\alpha_1^2 \geq \alpha_2^2 \geq \cdots \geq \alpha_K^2$, where $\alpha_k^2$ quantifies the SNR associated with the $k$-th component. The eigentasks are defined through the generalized eigenvalue problem
\begin{align}
    \mat{V}\vec{r}^{(k)} = \frac{1}{\alpha_k^2}\,\mat{G}\vec{r}^{(k)},
    \label{eq:eigentaskProb}
\end{align}
where $\mat{V}$ is the empirically estimated input-averaged readout-noise covariance matrix. Eigentasks depend on both the covariance matrix $\mat{V}$ and the Gram matrix $\mat{G}$, which encode the signal structure through mean responses of the sensor (see Appendix\,\ref{app:methods}). For eigentask-based noise mitigation, the transformation in \eqnref{eq:noisemit} is chosen as $\mat{W}_{\mathrm{EGT}}^{\mathsf{T}} = (\vec{r}^{(1)} \ \cdots \ \vec{r}^{(\Kr)})$, corresponding to projection onto the leading $\Kr$ eigentasks (see \figpanelref{fig:mnist_main}{a}). The transformation $\mat{W}_{\mathrm{EGT}}$ is estimated from training data and applied to measured features $\bar{\vec{X}}(\vec{u}^{(c)})$ to obtain eigentask features $\bar{\vec{Y}}_{\mathrm{EGT}}(\vec{u}^{(c)})$. This procedure is applicable across noise models, including Poissonian noise in photon counting and multinomial noise in measurements of qubit-based quantum systems.

We now outline the noise-mitigation procedure, which follows a common workflow across all tasks considered in this work. The first step is the construction of the transformation matrix $\mat{W}$ in \eqnref{eq:noisemit} (see \figpanelref{fig:schematic}{b}). For Fourier-domain low-pass filtering and spatial coarse graining, the transformation is data-agnostic, and $\mat{W}$ is fixed by the choice of output dimension $\Kr$. In contrast, PCA and eigentask learning are data-driven and require estimates derived from sensor measurements. For each input $\vec{u}$ in the training set, the averaged sensor output $\bar{\vec{X}}(\vec{u})$ is obtained by averaging over $S$ shots, up to a maximum of $S_\mathrm{max}$. Repeating this procedure over the training set yields a collection of measured features, from which the empirical covariance matrix $\mat{V}$ and Gram matrix $\mat{G}$ are estimated (see Appendix\,\ref{app:methods}). For PCA, $\mat{W}_{\mathrm{PCA}}$ is constructed from the eigenvectors of covariance matrix $\mat{C}$ across training inputs. For eigentask learning, $\mat{W}_{\mathrm{EGT}}$ is obtained by solving the generalized eigenvalue problem in \eqnref{eq:eigentaskProb} using both $\mat{V}$ and $\mat{G}$.

In practice, for both PCA and eigentask transformations, the basis $\mat{W}$ is constructed using the maximum available sampling budget, $S_\mathrm{max}$ shots per input. Inference performance is then evaluated under reduced sampling, $S < S_\mathrm{max}$, thereby isolating the impact of inference-time sampling noise. Conceptually, the number of shots used to construct the basis, $S_{\mathrm{basis}} = S_\mathrm{max}$, corresponds to a one-time calibration cost that can be amortized over subsequent inference tasks.

We emphasize that for all the noise-mitigation techniques, $\mat{W}$ does not need to be determined via an iterative optimization routine; instead, an explicit construction is known, parameterized only by $\Kr$, which functions as a hyperparameter. Once $\mat{W}$ is constructed for each noise-mitigation technique, it is applied - as defined by \eqnref{eq:noisemit} - to sensor data from the training set, and a logistic regression or neural network-based classifier $\mathcal{F}\{\cdot\}$ is trained. During the testing phase, the averaged sensor output is now computed for inputs $\vec{u}$ in the unseen testing set with $S=S_{\mathrm{infer}}\leq S_\mathrm{max}$ shots and test accuracy is computed from the inferences, the process of which captures the resource-constrained operating regime where fast, low-light measurements are required.

In what follows, we show that eigentask-based representation transforms frequently outperform or match other noise-mitigation approaches for a variety of machine-vision classification tasks, across high and low-SNR conditions and different physical systems with varying readout noise characteristics.

\section{\label{sec:result}Results}

\subsection{\label{subsec:result_mnist_lens}Eigentasks outperform other standard noise-mitigation methods: MNIST classification with a single-lens front end}

\begin{figure}[H]
    \centering
    \includegraphics[scale=\scale]{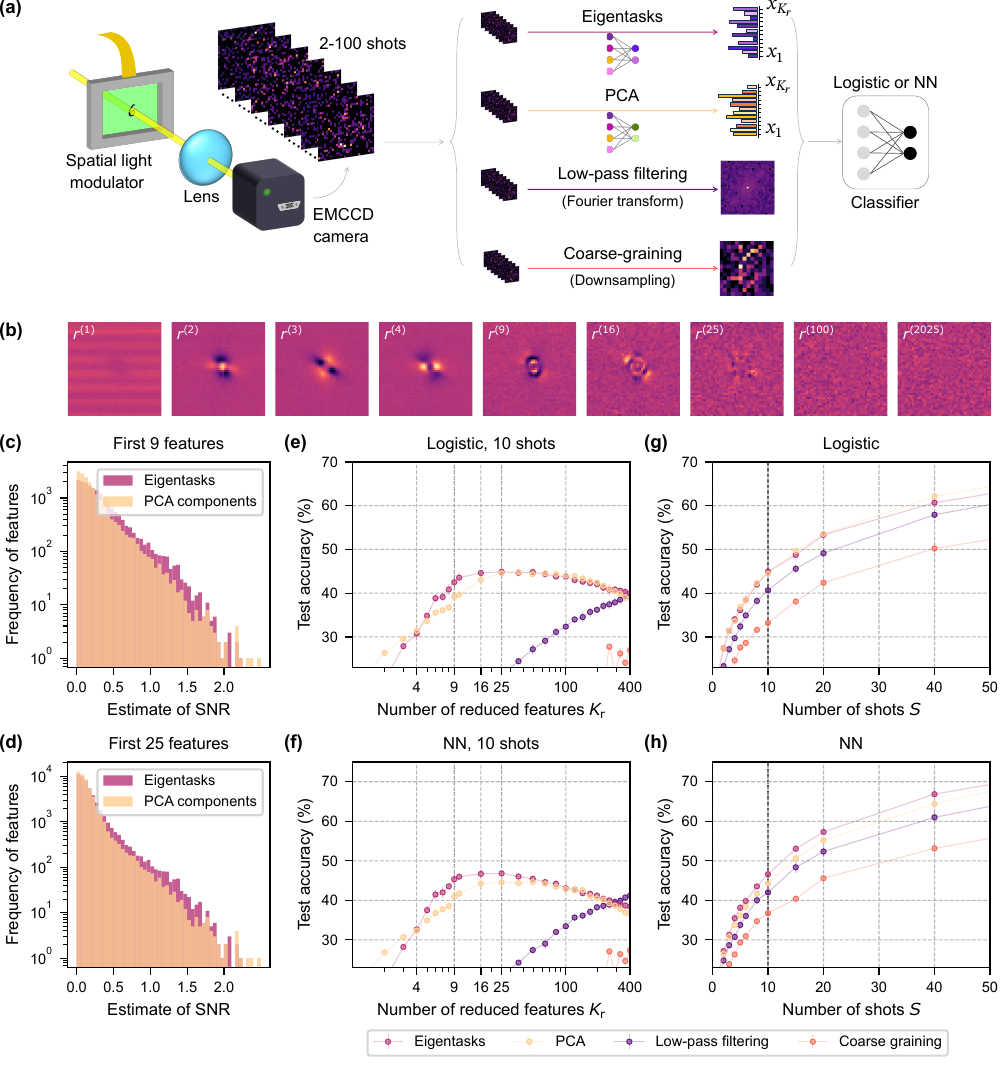}
    \caption{
    \figtitle{Comparative evaluation of noise-mitigation techniques for MNIST classification under photon-limited conditions using a simple lens-based optical front end.} 
    \figpanel{a} Experimental pipeline and evaluation workflow. An input MNIST digit is encoded in phase on an SLM, mapped by a fixed single-lens optical front end to an EMCCD feature vector $\vec{X}(\vec{u})$, averaged over $S$ shots, transformed to $\Kr$ reduced features, and classified. 
\figpanel{b} Examples of eigentask masks $\vec r^{(k)}$ (reshaped to the sensor grid) at low optical power; higher-order masks become noise-dominated.
    \figpanel{c}{d} Histograms of an empirical per-feature SNR (estimated from 100-shot data) for the first 9 or 25 retained PCA features and the leading 9 or 25 non-normalization eigentask features.
    \figpanel{e}{f} Test accuracy versus reduced dimension $\Kr$ at $S=10$ shots for logistic regression and neural network, respectively, comparing the four transforms. The two vertical dashed lines indicate the positions of 9 and 25 features in panel \figpanel{c} and \figpanel{d}. 
    \figpanel{g}{h} Test accuracy versus number of shots $S$ for logistic regression and neural network, respectively. The vertical dashed line indicates the position of $S=10$ shots plotted in panel \figpanel{e} and \figpanel{f}. 
    Model selection is described in Appendix\,\ref{app:methods}. Accuracy panels are means over five independent training/validation/test splits; error bars denote one standard deviation. Histograms and masks are shown for a representative split.
}
    \label{fig:mnist_main}
\end{figure}

We first examine machine-vision tasks using an SLM-lens-camera optical setup with a simple single-lens front end, illustrated in \figpanelref{fig:mnist_main}{a}. In this setup, a coherent incident beam passes through a spatial light modulator (SLM), which encodes the input $\vec{u}$—a grayscale image represented as a two-dimensional array of pixel values—by modulating the optical field phase through a mapping that links image pixels to the phase shifts applied by the SLM pixels. The beam then passes through a lens, and an EMCCD camera placed at the focal plane records the light intensity at each pixel; the $k$-th pixel value defines the sensor output $X_k$. Each shot captured by the EMCCD camera corresponds to a single shot. Further details of the experimental setup can be found in \suppnoteref{si:lens_setup}.

For this front end, we evaluate performance on the MNIST handwritten digits recognition task \cite{lecun2002gradient}, commonly used as a benchmark dataset for classification problems. We collected 100 shots per digit, with each shot obtained using a $45\times 45$ pixel grid ($K=45^2$), for a total of 12{,}000 digits: 8{,}000 training digits, 2{,}000 validation digits, and 2{,}000 test digits. Data were collected at two power levels: high power (approximately 696 detected photons per shot), and a significantly lower power (approximately 33 detected photons per shot, see Appendix\,\ref{app:methods} for the estimation of power) to enable an assessment of noise-mitigation techniques across varying signal-to-noise ratios. 

The eigentask coefficients $\vec{r}^{(k)}$ are obtained by solving the generalized eigenvalue problem from $S_\mathrm{max}=100$-shot data, and we visualize representative masks (reshaped to the $45\times 45$ camera grid) for low-power data in \figpanelref{fig:mnist_main}{b}. Additional masks are shown in \suppnoteref{si:eigentask_vis}. Applying the $k$-th mask to a shot-averaged readout yields the corresponding eigentask feature, $\bar{Y}_{k,\mathrm{EGT}}(\vec{u})=\vec{r}^{(k)\,\mathsf{T}}\bar{\vec{X}}(\vec{u})$. Consistent with the eigentask construction, low-order masks with large SNR exhibit clear spatial structure, whereas higher-order masks become progressively dominated by noise; in the low-power setting, the masks become visually noise-like around $k\approx 25$. The first set of weights, $\vec{r}^{(1)}$, is a near-uniform normalization vector, reflecting the approximately conserved total optical power.

To quantify how noise is redistributed by the learned bases, \figpanelref{fig:mnist_main}{c}{d} plot histograms of an empirical per-feature SNR estimate computed from the $S_\mathrm{max}=100$ shots (see Appendix\,\ref{app:methods} for the precise definition). Across both the first 9 and first 25 extracted non-normalization features, the eigentask features exhibit a heavier high-SNR tail than the corresponding PCA features, indicating that eigentask learning concentrates information into fewer, more reliably observable components. This SNR-aware ordering is central to the original eigentask framework, where the eigentask basis is explicitly optimized against sampling/readout noise rather than variance alone.

In \figpanelref{fig:mnist_main}{e}{f}, we evaluate MNIST classification at low optical power using features produced by the four different noise-mitigation techniques, sweeping the reduced feature dimension $\Kr$. Panel \figpanel{e} uses logistic regression with cross-entropy loss, and panel \figpanel{f} uses a multilayer perceptron classifier (see Appendix\,\ref{app:methods} for architectures and training details). In both panels, performance at each $\Kr$ is reported at the validation-selected epoch, and the results shown correspond to $S=10$ shots, with similar qualitative trends for other small-$S$ values. 

We observe a behavior that is characteristic of supervised learning in the finite-sample regime: increasing the number of features does not necessarily lead to saturation in performance and can instead degrade it~\cite{hughes1968mean, trunk1979problem, hua2005optimal}. This reflects overfitting during training. Adding features introduces additional noise; while controlled noise can act as a regularizer, incorporating features that carry little information leads the classifier to fit noise, resulting in degraded generalization on test data.

Notably, the maximum achievable accuracy and at what $\Kr$ it is achieved differs significantly across noise-mitigation techniques. In particular, eigentask-based representation attains its peak performance near $\Kr \simeq 25$, consistent with the onset of noise-dominated eigentask weights shown in \figpanelref{fig:mnist_main}{b}. This indicates that these methods differ not only in noise suppression but also in their alignment with informative structure in the data. Crucially, this limitation cannot be overcome by simply increasing the number of features. Consequently, the choice of noise-mitigation technique can fundamentally limit learning performance for a given noisy dataset.

For example, both spatial and Fourier-domain filtering techniques are suited to noise reduction, but are agnostic to useful structure in the data, which is fundamental for classification tasks. Such schemes are typically optimal for an intermediate number of features $\Kr$, where they reduce noise without completely losing all information present in the data. However, both schemes fare worse than the PCA and eigentasks. PCA is a technique built for dimensionality reduction and compression, and therefore achieves its maximum accuracy for a small $\Kr$, making use of the high compressibility and the substantial low-structure dimensional structure of image data in general~\cite{field1987relations, pope2020intrinsic, jolliffe2016principal}. However, the PCA is intended to maximize variance in a dataset to identify important components, without specifically taking note of SNR. Note that if our task was one of data compression only - representing a dataset with as few components as possible - the addition of more PCA components would always improve the representation. However, in supervised learning the addition of features leads to overfitting, as observed here.

The above schemes highlight the importance of eigentask learning as a noise-mitigation technique for supervised learning tasks: it enables the compression of data into a few information-rich components that are designed to maximize SNR, thereby being robust against overfitting during training. We therefore see that eigentasks outperform PCA, using fewer $\Kr$ to achieve similar or higher max accuracy.

To summarize performance as a function of sampling budget, we also select, for each method, the feature count $\Kr$ that maximizes validation accuracy and plot the resulting test accuracy versus shot count $S$ in \figpanelref{fig:mnist_main}{g}{h}.
As $S$ increases, sampling noise decreases and all methods improve, while eigentask learning remains consistently best or near-best, with the clearest gains in the few-shot regime.
We additionally compare raw versus thresholding preprocessing, which is often used in EMCCD photon-counting workflows to mitigate noise~\cite{basden2003photona, lantz2008multiimaging, harpsoe2012bayesian}, in \suppnoteref{si:thresholding_compare}, and find that eigentask learning remains robust under both preprocessing choices, whereas competing methods exhibit larger variability. High-power data are analyzed in \suppnoteref{si:mnist_high_power}. The results show qualitatively similar trends, indicating that the advantages of eigentask-based processing persist across different measurement conditions.

\subsection{Robustness of eigentask learning over different tasks: MPEG-7 classification with single-lens optical front end}

\begin{figure}[hbt]
    \centering
    \includegraphics[scale=\scale]{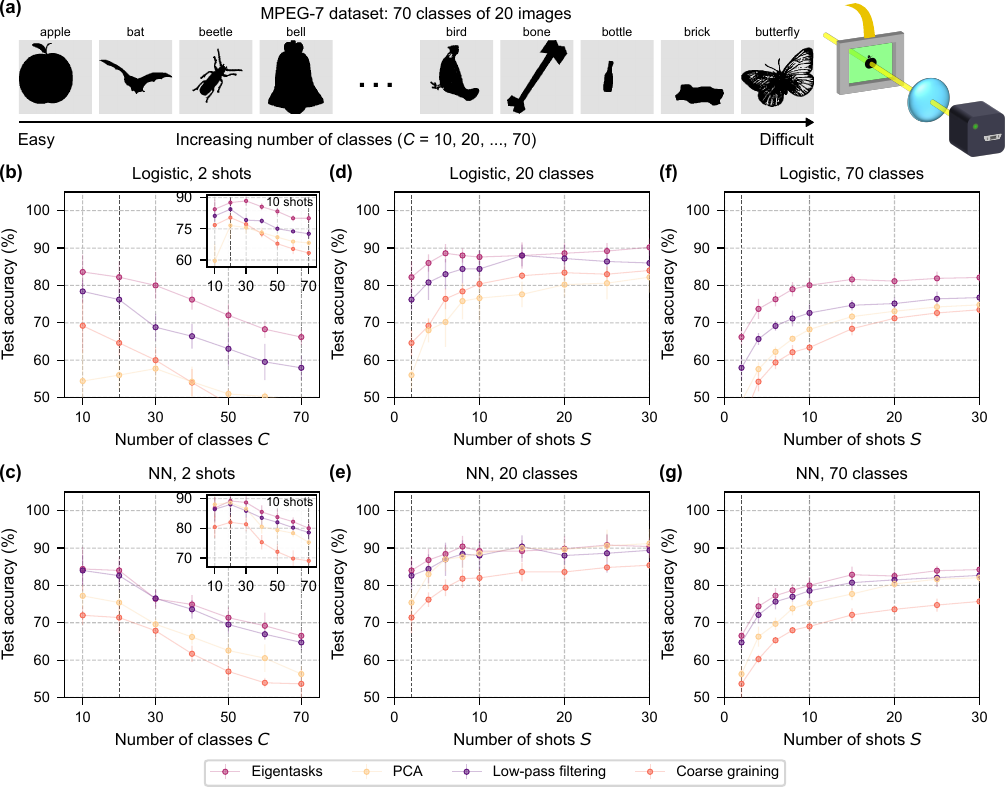}
    \caption{
    \figtitle{Comparative evaluation of noise-mitigation techniques for MPEG-7 classification under photon-limited conditions using a simple lens-based optical front end. }\figpanel{a} MPEG-7 classification task is performed using the single-lens sensing pipeline; task difficulty is tuned by changing the number of classes $C=10,20,\ldots,70$.
    \figpanel{b}{c} Test accuracy versus number of classes $C$ in the few-shot regime ($S=2$) using logistic regression and a neural-network classifier, respectively; insets show the corresponding trend for a higher-shot condition ($S=10$).
    \figpanel*{d}{g} Test accuracy versus number of shots $S$ for $C=20$ classes using \figpanel{d} logistic regression and \figpanel{e} a neural network, and for $C=70$ classes using \figpanel{f} logistic regression and \figpanel{g} a neural network.
    Model selection is described in Appendix\,\ref{app:methods}. Data are means over five independent training/validation/test splits; error bars denote one standard deviation.
}
    \label{fig:mpeg7_main}
\end{figure}

While MNIST is a standard benchmark, we find that eigentask-based processing yields an even clearer advantage in more challenging classification settings. We therefore consider few-shot classification on the MPEG-7 dataset~\cite{bober2001mpeg7}, which contains 70 classes with 20 images per class (\figpanelref{fig:mpeg7_main}{a}). Using the same single-lens sensing pipeline as in the MNIST experiment, we recorded up to $S_\mathrm{max}=300$ shots per image on a $39\times 39$ sensor grid ($K=39^2$), at an incident power corresponding to approximately 798 detected photons per shot. For each class, we held 5 images in the test set, while using the remaining 15 for model selection by three-fold cross-validation, with each fold using 10 for training and 5 for validation; after selecting parameters such as $\Kr$ and epochs, the model was retrained on all 15-non-test images and evaluated on the held-out test set, the results of which are reported below. Such procedures were repeated over five random training/validation/test splits (see Appendix\,\ref{app:methods}). In addition to varying the shot budget $S$, we tune task difficulty by increasing the number of classes $C$ from as few as 10, where the class count is comparable to MNIST, up to the full dataset of 70 classes. For a controlled comparison across class counts, we use a fixed nested ordering of classes, such that $C$-class task is formed from the first $C$ classes, with randomness introduced only through train/test splits.

\Figpanelref{fig:mpeg7_main}{b}{c} summarize the performance in the few-shot regime ($S=2$) as a function of $C$ for logistic regression \figpanel{b} and a neural-network classifier \figpanel{c}. For each method and each $(C,S)$ setting, we sweep the reduced feature dimension $\Kr$; for each $\Kr$, the epoch is selected by mean validation accuracy of the three folds, and the reported accuracy is then taken at the validation-selected $\Kr$ (see \figref{fig:mpeg7_main} caption and Appendix\,\ref{app:methods}). In this few-shot regime where sampling noise is dominant, eigentasks consistently provide the highest accuracy, with the largest gains emerging at larger $C$: in logistic regression, the eigentask advantage grows to \,$\sim$10 percentage points once $C\gtrsim 20$, while in the neural-network back end the improvement remains positive but more modest (typically a few percentage points), consistent with the reduced sensitivity of a more expressive classifier to imperfect feature representations.

As the number of shots increases, sampling noise is reduced, and all methods improve. The performance gap between the eigentasks and the other methods narrows. Nevertheless, eigentasks maintain an edge in the moderate-shot regime, as shown in the insets of \figpanelref{fig:mpeg7_main}{b}{c} at $S=10$. This trend is further confirmed in \figpanelref*{fig:mpeg7_main}{d}{g}, which plot accuracy versus $S$ for fixed class counts ($C=20$ and $C=70$) for both logistic regression and neural networks, with model selection procedure similar to \figpanelref{fig:mnist_main}{g}{h}. Overall, these results highlight that eigentask-based noise-mitigation remains robust across increasing task difficulty and shot budgets, and is especially beneficial in the few-shot, large-$C$ regime relevant to photon-limited sensing.

\subsection{\label{subsec:spdnn_results}Robustness of eigentask learning over different optical sensing architectures: MNIST classification with SPDNN encoder}

We have thus far analyzed the use of eigentasks for noise mitigation when applied to data obtained from simple optical sensors, which are not specifically optimized for supervised learning tasks. However, with increasing interest in trainable physical neural networks (PNNs)~\cite{wright2022deep, xue2024fully, momeni2025training}, optical sensors can be trained to maximize performance under constrained illumination and sampling resources. For such systems one may argue that the role of noise in optical measurement data may be reduced. By applying eigentask learning to data obtained from such an optimized stochastic optical encoder, we show that even in such applications there are still advantages to be extracted by noise-mitigation techniques.

\begin{figure}[!htbp]
    \centering
    \includegraphics[scale=\scale]{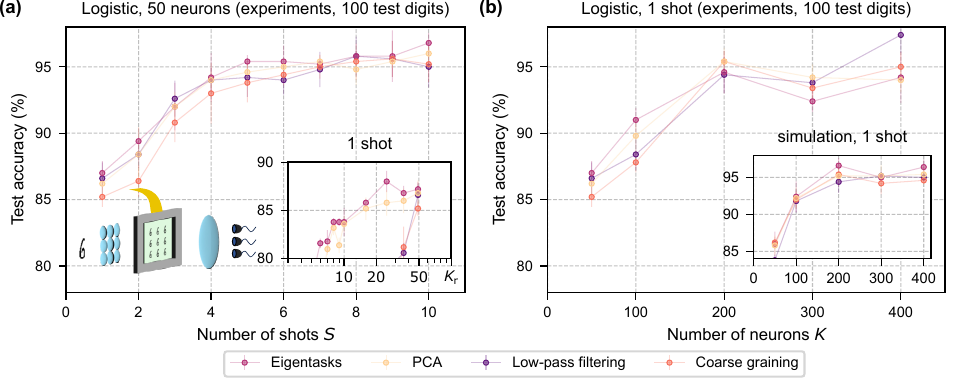}
    \caption{
    \figtitle{Performance of noise-mitigation techniques in MNIST classification using SPDNN experimental data (Ref.~\cite{ma2025quantumlimited}) with logistic regression.}
    \figpanel{a} Test accuracy on 100 test digits from experimental data (Ref.~\cite{ma2025quantumlimited}) versus number of shots $S$ for a 50-neuron SPDNN. The inset shows accuracy versus the reduced feature dimension $\Kr$ at $S=1$. \figpanel{b} Test accuracy on the same 100 digits versus neuron count $K$ for experimental data at $S=1$ shot, with an inset showing the corresponding results for simulated data on the same 100 digits. Model selection is described in Appendix\,\ref{app:methods}. Data are means over five independent simulated training/validation subsets; error bars denote one standard deviation.
See \suppfigref{sifig:spdnn_sim} for simulated performance on a larger test set.}
    \label{fig:spdnn_main}
\end{figure}

The trained task-specific optical encoder considered here is a one-layer single-photon-detection neural network (SPDNN); a schematic of this setup is shown in the inset of \figpanelref{fig:spdnn_main}{a} (see \suppnoteref{si:spdnn_setup} for details). An SPDNN is an optical neural network comprising an optical matrix-vector multiplier (MVM), which performs the linear operations of a neural network, and an array of single-photon detectors (SPDs), which implement stochastic binary activation functions. In this setup, the binary activation generated by the $k$-th SPD serves as the output feature $X_k$, forming the basis of the SPDNN's functionality. While being intrinsically extremely stochastic, Ma \textit{et al.}~\cite{ma2025quantumlimited} have shown that by including a faithful model of SPD activation in the training loop, they could perform deterministic classification with this kind of stochastic optical neural network (ONN). SPDNNs differ from the experimental setup studied before as the weights have been trained to maximize the probability of some activations or `clicks' given a specific input data. Thus it is \textit{a priori} unclear how applying a post-activation noise-mitigation procedure can improve the performance attained in Ref.~\cite{ma2025quantumlimited}.

Ma \textit{et al.} \cite{ma2025quantumlimited} demonstrated excellent agreement between simulated and experimental data, with training performed on simulated data and successfully validated on their experimental measurements. In this section, we do \emph{not} perform new SPDNN experiments; instead, we reanalyze the experimental SPDNN outputs reported in Ref.~\cite{ma2025quantumlimited} and compare against corresponding simulations generated using their validated model. We learn the eigentask and PCA bases and train the downstream logistic classifier, both using simulated SPDNN outputs. The performance is evaluated on both experimental data from Ref.~\cite{ma2025quantumlimited} and simulated test data. Networks with varying numbers of neurons—$K\!\!=$50, 100, 200, 300, and 400—are trained using simulated data of 8{,}000 digits with models selected based on another simulated 2{,}000 digits. Test results on 2{,}000 simulated digits, with single-shot data, show that eigentask learning achieves best or near-best performance across all methods (see \suppnoteref{si:spdnn_sim}). Consequently, we further test the method on experimental data of 100 digits, also with single-shot data, along with the corresponding simulations of these 100 digits.

\Figpanelref{fig:spdnn_main}{a} shows the test accuracy of the trained model on 100 experimental test digits as a function of the number of shots $S$ for the 50-neuron setup. Accuracy increases rapidly with $S$, exceeding 95\% with as few as $S=5$ shots, and eigentask learning remains best or near-best across the tested shot range. The inset depicts the accuracy as a function of reduced feature count $\Kr$ for single-shot data, where overfitting is observed, as noted previously. \Figpanelref{fig:spdnn_main}{b} presents the test accuracy as a function of neuron count $K$ for single-shot experimental data, with the inset giving the corresponding simulated results for the same 100 digits. Performance generally improves as the neuron count increases, although the trend is not strictly monotonic on the small experimental test set. The advantage of eigentask learning is most apparent in smaller neuron count regime, where it delivers highest or near-highest accuracy among the four methods. We also report results using a neural-network back-end classifier in \suppnoteref{si:spdnn_nn}.

Overall, the results in \figref{fig:spdnn_main} and the Supplementary Information indicate that eigentask learning performs robustly across both simple lens-based front ends and optimized SPDNN-based encoders in few-shot, noisy regimes, supporting its effectiveness across different architectures.

\section{Discussion}
\subsection{Summary of Results}
In this work, we demonstrate that eigentask learning provides an effective, measurement-adapted representation for optical image sensor readout, particularly when operating under constrained photon budgets and measurement noise. By ordering features according to their resolvability under noise rather than variance alone, eigentasks often extract more informative low-dimensional representations. For the MNIST dataset using a simple single-lens front end at low optical power, eigentask representation optimized the feature space such that peak classification accuracy was achieved with just $25$ reduced features (lower than other noise mitigation techniques considered), before the onset of noise-dominated overfitting. The performance enhancement became more pronounced as task complexity increased: in the 70-class MPEG-7 few-shot classification ($S=2$), eigentask representation outperformed baseline methods like PCA by approximately 10 percentage points when the class count exceeded 20 ($C \gtrsim 20$) using a logistic regression classifier. Furthermore, applying eigentasks to an optimized single-photon-detection neural network (SPDNN) yielded an increase in sample efficiency; the 50-neuron experimental setup reached approximately 95\% accuracy on the 100-digit experimental test set using only 5 shots. In broader simulated validations of 2{,}000 digits, eigentasks maintained a consistent best or near-best performance.

These results indicate that performance in photon-limited optical systems is not determined solely by the optical front end. When a limited photon budget constrains measurement reliability, gains can be achieved by adapting how the readout is represented prior to inference. This complements existing approaches based on optical design and suggests that representation deserves to be treated as a central element of the sensing pipeline alongside optical design. The scope of the present results is limited to regimes in which noise and data constraints are significant. As the number of measurements increases, the performance gap between methods narrows. In addition, sufficiently expressive downstream models may approximate similar transformations given ample data. The advantage of eigentasks therefore lies in providing an explicit, measurement-adapted representation that improves robustness and sample efficiency when data are limited.

\subsection{Limitations}
Despite its advantages in noise-dominated environments, this approach exhibits specific operational and algorithmic constraints. First, the construction of the eigentask basis requires a one-time calibration cost, utilizing a maximum sampling budget ($S_\mathrm{max}$) to accurately estimate the empirical covariance and Gram matrices from training data. Additionally, the relative advantage of eigentask learning diminishes as the signal-to-noise ratio increases. At higher optical powers or larger shot counts, the performance of standard baseline methods converges with that of eigentasks. We also observed that more expressive, nonlinear downstream classifiers (such as multilayer neural networks) can partially compensate for unoptimized feature spaces, narrowing the performance gap between eigentasks and other methods. For example, in the SPDNN 2{,}000-digit simulated test using a neural network, PCA surpassed eigentasks in the extreme few-shot regime ($S=1\text{--}3$). Finally, similar to other feature-extraction methods in finite-sample supervised learning, adding too many eigentask features eventually introduces noise that degrades generalization, highlighting the necessity of careful dimensionality tuning.

\subsection{Outlook}
A promising future direction is the direct integration of eigentask learning into the training loop of physical neural networks (PNNs), moving beyond its current implementation as a post-readout preprocessing step. Because eigentasks are model-free and driven by empirical data, they are naturally compatible with iterative optimization routines. This presents an opportunity for a co-designed pipeline where the sensor, its measurement representation, and the digital algorithm back end are jointly optimized~\cite{zhou2011computational, mait2018computational}. Such end-to-end architectures may significantly improve sample efficiency, paving the way for more resilient optical inference in applications constrained by latency and photon budgets.

\begin{acknowledgments}

This material is based upon work supported by the Air Force Office of Scientific Research under award number FA9550-22-1-0203. Research was sponsored by the Army Research Office and was accomplished under Grant Number W911NF-25-1-0261. The views and conclusions contained in this document are those of the authors and should not be interpreted as representing the official policies, either expressed or implied, of the Army Research Office or the U.S. Government. The U.S. Government is authorized to reproduce and distribute reprints for Government purposes notwithstanding any copyright notation herein.

The authors declare no competing interests.

\end{acknowledgments}

\section*{Data and code availability}

The experimental lens-based datasets (low-power MNIST, high-power MNIST, and MPEG-7), the simulated and experimental SPDNN MNIST outputs, and the precomputed training results supporting the figures of this study are available at Zenodo (\url{https://doi.org/10.5281/zenodo.19888614})~\cite{chen2026eigentask_zenodo}. Study-generated data and precomputed results are released under a CC BY 4.0 license, with third-party dataset components remaining subject to their original license terms. 
The code used for training, analysis, EMCCD calibration and figure generation is available at \url{https://github.com/TomCty1120/optical-eigentask-learning} under an MIT license.

\appendix

\section{\label{app:methods}Methods}

\subsection{EMCCD gain calibration and photon-budget estimation}

For the lens-based experiments, the EMCCD gain was calibrated from the shot-to-shot fluctuations of the raw camera readout.  In the large-gain limit, the photon-induced pixel values $X_\mathrm{ph}$ satisfy~\cite{hirsch2013stochastic, basden2003photona, lantz2008multiimaging}:
\begin{equation}
    \mean[X_\mathrm{ph}] = \eta\lambda g,\quad \Var[X_\mathrm{ph}] = 2\eta\lambda g^2,
\end{equation}
where $\lambda$ is the mean incident photon number during one exposure, $\eta$ is the quantum efficiency, and $g$ is the EMCCD gain; the mean and variance are taken across repeated shots. The measured raw EMCCD output is modeled as the sum of the photon-induced signal and a power-independent dark contribution, $X=X_\mathrm{ph}+X_\mathrm{dark}$. Matched dark shots, acquired with the same exposure time and camera settings but without incident light, were used to estimate the empirical dark mean $\bar X_\mathrm{dark}$.

Assuming the dark contribution is statistically independent of the photon-induced signal, 
\begin{equation}
    \mean[X] = \eta\lambda g+\bar{X}_\mathrm{dark},\quad \Var[X] = 2\eta\lambda g^2+\Var[X_\mathrm{dark}].
\end{equation}
Thus, when the shot-to-shot variance is plotted against the background-subtracted mean pixel value, the expected slope is $2g$. To estimate this slope, we grouped repeated measurements of the same input and camera pixel according to their shot-averaged pixel values. For each mean bin $m$, we computed the average within-trace shot variance $v$ and fitted
\begin{equation}
    \hat{v}_j=a(\hat{m}_j-\bar{X}_\mathrm{dark})+b,
\end{equation}
where $\hat m_j$ and $\hat v_j$ are the mean pixel value and shot-to-shot variance for bin $j$. The EMCCD gain was then estimated as $g=a/2$. Additional details of binning, quality control, and bootstrap uncertainty estimation are provided in \suppnoteref{si:lens_noise}.

The detected photon budget per input and per shot was
estimated from the non-thresholded, shot-averaged data as
\begin{equation}
    N_\mathrm{ph} = \frac{1}{gN_\mathrm{input}} \sum_{n=1}^{N_\mathrm{input}} \sum_{k=1}^{K} \big(\bar{X}_k(\vec{u}_n) - \bar{X}_\mathrm{dark}\big),
\end{equation}
where $N_\mathrm{input}$ is the number of input images and $K$ is the number of pixels per shot. The total incident photon budget is given by $N_\mathrm{ph}/\eta$.

\subsection{Estimation of eigentasks and the transformation matrix of eigentask learning}

For a specific input $\vec{u}$, the covariances are given by the covariance matrix $\mat{\Sigma}(\vec{u})\in \mathbb{R}^{K\times K}$ with $\mat{\Sigma}_{kk'}=\Cov_{\vec{\mathcal{X}}}[{\zeta}_k(\vec{u}),{\zeta}_{k'}(\vec{u})]=\mean_{\vec{\mathcal{X}}}[{\zeta}_k(\vec{u}){\zeta}_{k'}(\vec{u})]$. The covariance matrix is defined by taking the expectation value over the inputs $\mat{V}=\mean_{\vec{u}}[\mat{\Sigma}(\vec{u})]$. The Gram matrix is defined by $\mat{G}=\mean_{\vec{u}}[\vec{x}(\vec{u})\vec{x}(\vec{u})^\mathsf{T}]$.

In experiments, we collected outputs for $N_\mathrm{input}$ inputs from the dataset $\mathcal{D}$ and for each input $S_\mathrm{max}$ independently and identically distributed random outputs $\vec{\mathcal{X}}(\vec{u})=\{\vec{X}^{(s)}(\vec{u})\}_{s\in [S_\mathrm{max}]}$ were collected. We only use the $N_\mathrm{train}$, the size of training set, of them, which constitute the training set, to estimate the covariance and Gram matrices. With Bessel's correction considered, a finite-sample estimation of covariance matrix from the full record of single-shot data is given by:
\begin{equation}
    (\tilde{\mat{V}}_{N_\mathrm{train}})_{kk'}=\frac{1}{N_\mathrm{train}(S_\mathrm{max}-1)}\sum_{n=1}^{N_\mathrm{train}}\sum_{s=1}^{S_\mathrm{max}}(X_k^{(s)}(\vec{u}_n)-\bar{X}_k(\vec{u}_n))(X_{k'}^{(s)}(\vec{u}_n)-\bar{X}_{k'}(\vec{u}_n)),
\end{equation}
where $\bar{X}_k(\vec{u}_n)=\sum_{s=1}^{S_\mathrm{max}}X_k^{(s)}(\vec{u}_n)/S_\mathrm{max}$ is the estimated expectation value of the $k$-th feature, and the finite-sample estimation of the Gram matrix can be obtained by:
\begin{equation}
    (\tilde{\mat{G}}_{N_\mathrm{train}})_{kk'}=\frac{1}{N_\mathrm{train}}\sum_{n=1}^{N_\mathrm{train}}\bar{X}_{k}(\vec{u}_n)\bar{X}_{k'}(\vec{u}_n).
\end{equation}
In the large-$N_\mathrm{train}$ limit, the covariance matrix is unbiased $\mat{V}=\tilde{\mat{V}}\equiv\lim_{N_\mathrm{train}\rightarrow \infty}(\tilde{\mat{V}}_{N_\mathrm{train}})$, while the Gram matrix needs to be corrected according to:
\begin{equation}
\begin{aligned}
    \tilde{\mat{G}}_{kk'}&\equiv\lim_{N_\mathrm{train}\rightarrow\infty}(\tilde{\mat{G}}_{N_\mathrm{train}})_{kk'}=\mean_{\vec{u}}[\mean_{\vec{\mathcal{X}}}[\bar{X}_k(\vec{u})\bar{X}_{k'}(\vec{u})]]
    \\&=\mean_{\vec{u}}\qty[\mean_{\vec{\mathcal{X}}}\qty[\qty(x_k(\vec{u})+\frac{1}{\sqrt{S_\mathrm{max}}}\zeta_k(\vec{u}))\qty(x_{k'}(\vec{u})+\frac{1}{\sqrt{S_\mathrm{max}}}\zeta_{k'}(\vec{u}))]]
    \\&=\mat{G}_{kk'}+\frac{1}{S_\mathrm{max}}\mat{V}_{kk'},
\end{aligned}
\end{equation}
and thus $\mat{G}=\tilde{\mat{G}}-\frac{1}{S_\mathrm{max}}\tilde{\mat{V}}$. We use the $\tilde{\mat{G}}_{N_\mathrm{train}}=\mat{G}+\frac{1}{S_\mathrm{max}}\mat{V}$ and $\tilde{\mat{V}}_{N_\mathrm{train}}=\mat{V}$ to estimate eigentasks by solving the generalized eigenvalue problem $\tilde{\mat{V}}_{N_\mathrm{train}}\tilde{\vec{r}}^{(k)}=\frac{1}{\tilde{\alpha}_k^2}\tilde{\mat{G}}_{N_\mathrm{train}}\tilde{\vec{r}}^{(k)}$. By comparing with \eqnref{eq:eigentaskProb}, we get the empirical estimation of eigentasks $\vec{r}^{(k)}=\tilde{\vec{r}}^{(k)}$, SNRs $\alpha_k^2=\tilde{\alpha}_k^2-\frac{1}{S_\mathrm{max}}$, and the transformation matrix is $\mat{W}_\mathrm{EGT}^\mathsf{T}=(\vec{r}^{(1)}\cdots\vec{r}^{(\Kr)})$ and the bias is $\vec{b}=\vec{0}$. This transformation matrix will be applied not only to $S_\mathrm{max}$-shot averaged data but also fewer shots $S=S_{\mathrm{infer}}\leq S_\mathrm{max}$.

\subsection{Estimation of principal components and the transformation matrix of PCA}

Similar to eigentasks, we also estimate principal components and the transformation based on the training set with the maximum number of shots $S_\mathrm{max}$ collected and apply them to the full dataset with fewer shots. 

Firstly, the $S_\mathrm{max}$-shot mean of features $\bar{\vec{X}}(\vec{u})$ are calculated and the mean vector of features estimated from training set is subtracted to obtain zero-centered features $\bar{\vec{Z}}(\vec{u})=\bar{\vec{X}}(\vec{u})-\frac{1}{N_\mathrm{train}}\sum_{n=1}^{N_\mathrm{train}}\bar{\vec{X}}(\vec{u}_n)$. All the following processing is with respect to the zero-centered data. Afterwards, a matrix $\mat{C}$ is calculated as:
\begin{equation}
    \mat{C}_{kk'}=\sum_{n=1}^{N_\mathrm{train}}\bar{Z}_k(\vec{u}_n)\bar{Z}_{k'}(\vec{u}_n).
\end{equation}
Considering the zero-centering step, the $\mat{C}$ matrix is essentially the covariance matrix of mean features across different inputs. The principal components are the eigenvectors $\vec{v}^{(k)}$ of $\mat{C}$ satisfying $\mat{C}\vec{v}^{(k)}=\sigma_k^2\vec{v}^{(k)}$ with descending order of variance eigenvalues $\sigma_1^2\geq\sigma_2^2\geq\cdots\geq\sigma_K^2$. All zero-centered data $\bar{\vec{Z}}(\vec{u})$ is projected to these principal components before being fed into $\mathcal{F}$ for prediction, yielding the transformation matrix $\mat{W}_\mathrm{PCA}^\mathsf{T}=(\vec{v}^{(1)}\cdots\vec{v}^{(\Kr)})$ and the bias $\vec{b}=-\mat{W}_\mathrm{PCA}(\frac{1}{N_\mathrm{train}}\sum_{n=1}^{N_\mathrm{train}}\bar{\vec{X}}(\vec{u}_n))$.

\subsection{Fourier-domain low-pass filtering and spatial coarse graining}

In Fourier-domain low-pass filtering, we apply the Fourier transform to the averaged features $\bar{\vec{X}}(\vec{u})$ and extract the real and imaginary parts of the low-frequency components as input to the output layer. For output features from EMCCD two-dimensional pixel arrays, the features can be expressed as $\bar{X}_k=\bar{X}_{i,j}$, where $i$ and $j$ are indices for the two dimensions, ranging from $-\lfloor \frac{L-1}{2}\rfloor$ to $\lfloor \frac{L}{2}\rfloor$, and $L$ is the side length of the array. The Fourier transform is applied to generate frequency components $f_{i,j}$. Since the frequency components satisfy $f_{i,j} = f_{-i,-j}^*$ (due to the real-valued nature of the features), we select the real and imaginary parts, $\Re[f_{i,j}]$ and $\Im[f_{i,j}]$, as input to the output layer. Specifically, we include components with $0 \leq i \leq \Lr/2$ and $-\Lr/2 \leq j \leq \Lr/2$, where $\Lr = \sqrt{\Kr}$ is the reduced side length. 

For the SPDNN data, the output features do not have an inherent two-dimensional spatial arrangement. Accordingly, the features are represented as a one-dimensional array, and a one-dimensional Fourier transform is applied. The reduced input to the output layer is then formed from the real and imaginary parts of the retained low-frequency Fourier components, following the same principle as in the two-dimensional case.

In 2D spatial coarse graining, $K=L\times L$ features are resized to $\Kr=\Lr\times \Lr$ features, corresponding to EMCCD data; in 1D spatial coarse graining, $K=L$ features are resized to $\Kr=\Lr$, corresponding to SPDNN data. Both coarse graining processes use linear interpolation. 

\subsection{Dataset-specific data splits and evaluation protocols}

For each input image $\vec{u}^{(n)}$, we collected up to $S_\mathrm{max}$ repeated stochastic measurements
$\{\vec{X}^{(s)}(\vec{u}^{(n)})\}_{s=1}^{S_\mathrm{max}}$, where $\vec{X}^{(s)} \in \mathbb{R}^{K}$ denotes the raw sensor readout from a single shot.
To emulate operation on a smaller sampling budget $S \le S_\mathrm{max}$, we formed the $S$-shot feature vector
\begin{equation}
\bar{\vec{X}}(\vec{u}^{(n)}) = \frac{1}{S}\sum_{s=1}^{S} \vec{X}^{(s)}(\vec{u}^{(n)}),
\end{equation}
using the first $S$ shots acquired for that input. For the MNIST lens-front-end experiments, $S_\mathrm{max}=100$ and $K=45^2$;
for the MPEG-7 lens-front-end experiments, $S_\mathrm{max}=300$ and $K=39^2$;
for the SPDNN data, $S_\mathrm{max}=30$ and $K\in\{50, 100, 200, 300, 400\}$ denotes the number of detector outputs. When thresholded EMCCD data were considered, thresholding was applied to every raw shot prior to shot averaging. 

For each independent repetition, the data were partitioned into training, validation, and test subsets before learning the noise-mitigation bases and downstream classifiers. For the MNIST and MPEG-7 lens-front-end experiments, class-balanced training, validation, and test subsets were formed from the measured data before learning the noise-mitigation transforms and downstream classifiers. In the SPDNN analysis, by contrast, training and validation data were taken from simulated outputs, whereas testing was performed on either experimental or simulated outputs. 

In the MNIST lens-front-end experiments, 12{,}000 measured digits were split into 8{,}000 training, 2{,}000 validation and 2{,}000 test samples, with class-balanced subsets. In the MPEG-7 lens-front-end experiments, each $C$-class task contained 20 images per class, from which 5 randomly chosen images were held for testing, and the remaining 15 images per class were used for training and model selection by three-fold cross-validation; for each class count $C$, we used a fixed nested subset consisting of the first $C$ classes in a predefined ordering, with randomness arising only from within-class training/validation/test split. In the SPDNN analysis, eigentask and PCA bases, as well as the downstream classifiers for all four methods, were learned from simulated training outputs of 8{,}000 class-balanced digits, evaluated and selected based on 2{,}000 class-balanced validation digits, and then tested on 100 experimental digits reported in Ref.~\cite{ma2025quantumlimited}, the corresponding simulated outputs of the same digits (see Section\,\ref{subsec:spdnn_results}), and a larger test set of 2{,}000 test digits used in \suppnoteref{si:spdnn_sim}. All training and testing procedures were repeated over five independent random seeds. For MNIST and MPEG-7 lens-based data, the repetitions used different training/validation/test splits. For SPDNN analysis, the simulated training and validation subsets were independently resampled for each seed. The 100-digit test set, either experimental or simulated, was fixed across seeds. In the larger 2{,}000-digit simulation-only analysis, the simulated test subset was also independently resampled for each seed.

\subsection{Downstream classifiers, optimization, and reporting}

For each $\Kr$, features were normalized using statistics computed from the training subset only. For eigentask and PCA, we first computed the full ordered set of transformed features $\bar{\vec{Y}}$, normalized these features, and then retained the leading $\Kr$ components. For low-pass filtering, the full Fourier-transformed features were normalized before retaining the low-frequency components. For coarse graining, the unreduced raw features were normalized before spatial downsampling. For the reported logistic-regression results, eigentask features used the non-zero-centered global rescaling, whereas PCA features, and the unreduced representations used for low-pass filtering and coarse-graining were feature-wise standardized. For neural-network classifiers, standardized features were used throughout. In feature-wise standardization, the training-set mean is subtracted and the result is divided by the training-set standard deviation for each feature; in global rescaling, a single training-set scale factor is used without zero-centering. Such selection was based on the performance comparison between different preprocessing, either standardization or global rescaling.

For downstream classification $\mathcal{F}\{\cdot\}$, we used either a multinomial logistic-regression classifier or a multilayer perceptron. The logistic-regression model consisted of a single linear layer mapping the $\Kr$-dimensional transformed features to $C$ output logits and was trained with cross-entropy loss. The neural-network classifier was implemented as a fully connected network with size $\Kr\times 400\times C$, i.e., one hidden layer of $400$ neurons, with a ReLU activation, without batch normalization. Both models were trained using AdamW optimizer with $\beta_1=0.9$ and $\beta_2=0.999$, zero weight decay, batch size $100$, and $300$ epochs. A ReduceLROnPlateau learning rate scheduler was used with a reduction factor $0.5$, patience of $10$ epochs, and minimum learning rate $10^{-5}$ for the MNIST dataset (both lens-based and SPDNN-based), and a StepLR learning rate scheduler was used for MPEG-7 dataset, reducing the learning rate by a multiplicative factor of $0.4$ every $T=50$ epochs.  The initial learning rate was $10^{-3}$ for all neural-network runs, and $10^{-3}$ or $0.5$ for logistic regression with standardized data or rescaled data, respectively. Random seeds were fixed for NumPy, Python, and PyTorch (including CUDA), and deterministic cuDNN settings were used.

For evaluation and reporting, validation and test accuracy were tracked for every epoch and every tested value of reduced feature count $\Kr$. In figures of accuracy versus reduced feature count $\Kr$, the reported value for each method corresponds to the test accuracy at epoch selected by validation accuracy for that $\Kr$. In figures of accuracy versus shot number $S$, class count $C$, or neuron count $K$, we further selected the reduced feature count using validation performance and reported the corresponding test accuracy. For the MNIST dataset, with lens front end or SPDNNs, this selection was performed directly on the held-out validation set. For MPEG-7 with limited data size, the selection was based on the mean validation accuracy across the three inner folds: within each fold, 10 images per class were used for training and 5 for validation. Ties were broken by preferring smaller $\Kr$, and then earlier epochs. After selecting hyperparameters, the classifier was retrained on all 15 non-test images per class and evaluated on the test set.

For low-pass filtering and coarse graining, we swept $\Kr=\Lr^2$ over all square dimensions for the two-dimensional EMCCD data and used the same numerical sequence of reduced feature counts for the one-dimensional SPDNN outputs; for eigentasks and PCA, we swept $\Kr\in\{1,2,3,\dots,10, 16, 25, 36,\dots\}$ up to the maximum dimension allowed by the feature space or the training-set size.

\subsection{Estimation of empirical per-feature SNRs}

After learning the transform with matrix $\mat{W}$ and bias $\vec{b}$ from the training set, we quantified how noise is redistributed by the transformed features using an empirical per-feature SNR estimate. For each input $\vec{u}$ in the test set, we applied the learned transform to each single-shot readout $\vec{X}^{(s)}(\vec{u})$ to obtain the noise-mitigated single-shot features $\vec{Y}^{(s)}(\vec{u})\equiv \mat{W}\vec{X}^{(s)}(\vec{u})+\vec{b}~(s=1,2,\dots,S)$. For each feature $Y_k(\vec{u})$, we then estimate its empirical SNR by taking the ratio between its mean $\bar{Y}_k(\vec{u})$ and its standard deviation $\delta Y_k(\vec{u})$ estimated from $S=S_\mathrm{max}$ samples (for MNIST data, $S_\mathrm{max}=100$):
\begin{align}
    \bar{\vec{Y}}(\vec{u})=\frac{1}{S_\mathrm{max}}\sum_{s=1}^{S_\mathrm{max}}\vec{Y}^{(s)}(\vec{u}),\quad \delta Y_k=\sqrt{\frac{1}{S_\mathrm{max}-1}\sum_{s=1}^{S_\mathrm{max}}({Y}_k^{(s)}(\vec{u})-\bar{Y}_k(\vec{u}))^2}\quad\implies\quad \snr[Y_k(\vec{u})]=\abs{\bar{Y}_k(\vec{u})/\delta Y_k(\vec{u})}.
\end{align}
By such estimation, we are able to obtain $\Kr N_\mathrm{test}$ empirical per-feature SNRs for noise-mitigated features. Histograms in \figpanelref{fig:mnist_main}{c}{d} and \suppfigpanelref{sifig:mnist_high_power}{a}{b} were formed by pooling these estimates over test inputs and over the selected leading features of $\Kr=9$ and $\Kr=25$. For eigentasks, the histogram excludes the leading component computed from $\vec{r}^{(1)}$ (which is a normalization component, see \suppnoteref{si:eigentask_vis}) and uses $\vec{r}^{(2)}$ onward. Standardization (see above) was applied.

\putbib[bibfile]
\end{bibunit}

\clearpage

\appendix

\begin{bibunit}[apsrev4-2]
\nocite{apsrev42Control}

\clearpage
\pagenumbering{arabic}
\setcounter{page}{1}

\setcounter{section}{0}
\setcounter{subsection}{0}
\setcounter{figure}{0}
\setcounter{table}{0}
\setcounter{equation}{0}

\renewcommand{\appendixname}{Supplemental Note}
\renewcommand{\thesection}{\arabic{section}}
\renewcommand{\thesubsection}{\Alph{subsection}}
\renewcommand{\thefigure}{S\arabic{figure}}
\renewcommand{\thetable}{S\arabic{table}}
\counterwithout{equation}{section}
\renewcommand{\theequation}{S\arabic{equation}}

\begin{center}
{\large\bfseries Supplemental Material for\\[0.35em]
``Measurement-Adapted Eigentask Representations for Photon-Limited Optical Readout''\par}
\vspace{0.75em}
Tianyang Chen, Mandar M. Sohoni, Saeed A. Khan, J\'er\'emie Laydevant,\\
Shi-Yuan Ma, Tianyu Wang, Peter L. McMahon, and Hakan E. T\"ureci
\end{center}

\vspace{1em}

\startcontents[Supplementary Note]
\printcontents[Supplementary Note]{l}{1}{\section*{Contents}\setcounter{tocdepth}{2}}

\newpage

\section{\label{si:lens_encoder}Details of lens-based front end}
\subsection{\label{si:lens_setup}Experimental setup of lens-based front end}
\begin{figure}[hbt]
    \centering
    \includegraphics[width=\textwidth]{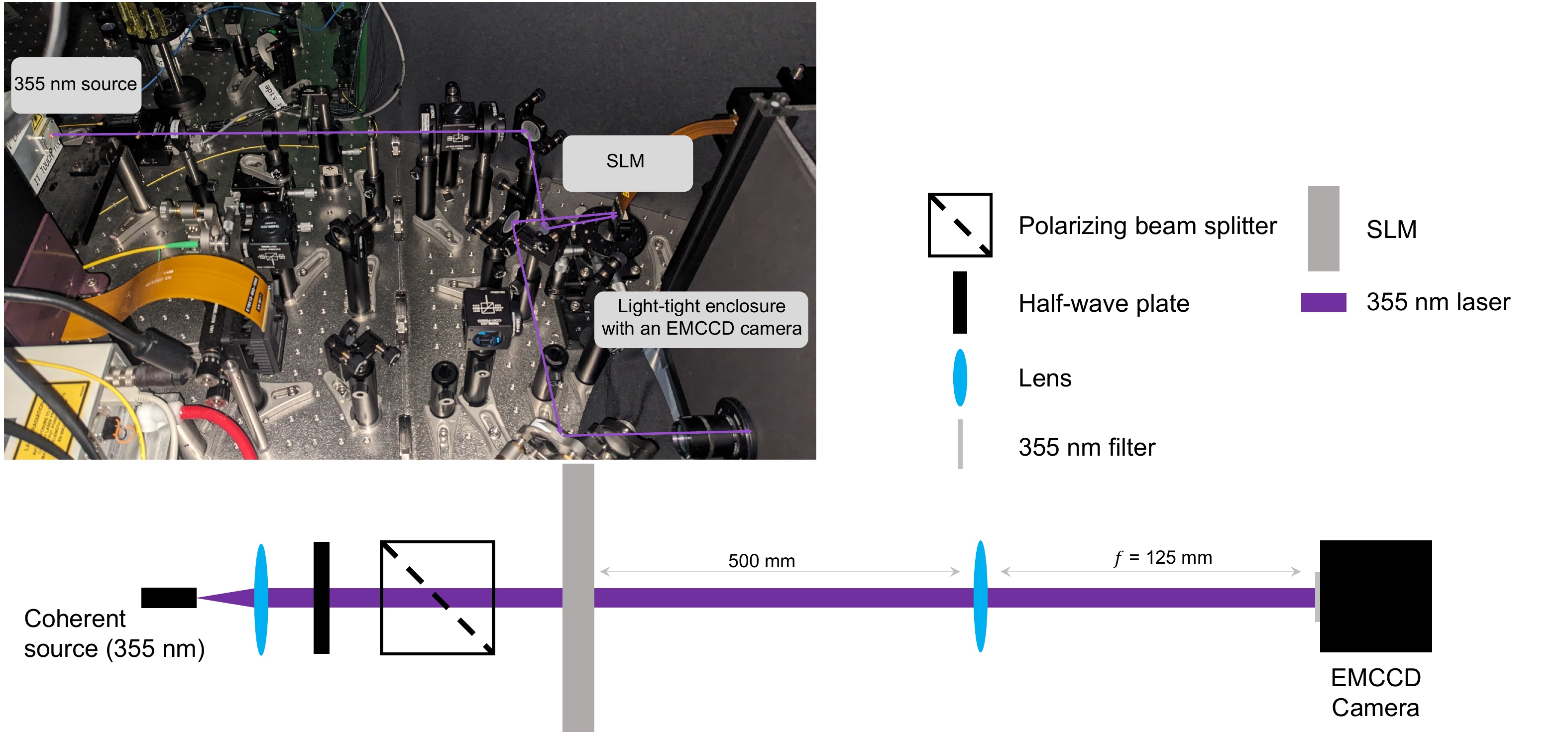}
    \caption{
    \figtitle{Schematic and picture of experimental setup.} An overview of the setup used to collect experimental data.}
    \label{sifig:experimental_setup}
\end{figure}

The SLM-lens-camera setup consisted of a \SI{355}{nm} laser (Xcyte, CY-355-0210M1), a spatial light modulator (SLM) (Holoeye, PLUTO-2.1-UV-099), single plano-convex lens ($f$ = \SI{125}{mm}, Thorlabs LA4236-UV-ML), and a Princeton Instruments electron-multiplying charge coupled device (EMCCD) camera (ProEM-HS: 512BX3). The MNIST~\cite{lecun2002gradient} and MPEG-7~\cite{bober2001mpeg7} dataset objects were displayed on the SLM and then imaged onto the camera using the plano-convex lens. The distance between the SLM and lens was about \SI{500}{mm}, and the distance between the camera and the lens was \SI{125}{mm}. \Suppfigref{sifig:experimental_setup} depicts the setup. 

\subsection{\label{si:lens_transform}Optical encoding map implemented by the lens-based optical front end and EMCCD readout}

A phase-only SLM does not by itself produce an observable intensity pattern; its modulation becomes visible only after subsequent optical propagation. In the current Fourier/focal-plane configuration~\cite{goodman1969introduction} used here, this role is played by the lens, which maps the phase-encoded field to a spatially varying intensity distribution at the camera plane, which the EMCCD measures. In the infinite-shot limit, the SLM-lens-camera pipeline realizes a deterministic optical encoding map, which we briefly analyze below.

The incident field is approximated as a Gaussian beam with complex field envelope
\begin{align}
    E_\mathrm{in}(x,y)=E_0 \exp\!\left[-\frac{x^2+y^2}{2w^2}\right],
\end{align}
where $w$ is the beam waist. The input $\vec{u}$ is reshaped into a two-dimensional array and encoded onto the phase-only SLM through a phase pattern $\phi_{\vec{u}}(x,y)$, taken here to be linearly proportional to the input,
\begin{align}
    \phi_{\vec{u}}(x,y)=\gamma_\phi\,u(x,y),
\end{align}
so that the encoded field becomes
\begin{align}
    E_\mathrm{enc}(x,y;\vec{u})
    =
    E_\mathrm{in}(x,y)\,e^{i\phi_{\vec{u}}(x,y)},
\end{align}
where $\gamma_\phi$ is the linear coefficient of the encoding. After propagation through the lens-based optical system, the field at the camera plane is
\begin{align}
    E_\mathrm{cam}(x,y;\vec{u})
    =
    \mathcal{P}\!\left[E_\mathrm{enc}(x,y;\vec{u})\right],
\end{align}
where $\mathcal{P}$ denotes the overall linear propagation operator from the SLM plane to the camera plane, close to a Fourier transform if passing through a lens imaging system.

The corresponding mean sensor readout is obtained by integrating the optical intensity over camera pixels. For the $k$-th pixel (or readout feature),
\begin{align}
    x_k(\vec{u})
    \propto
    \int_{\mathrm{pixel}\,k}
    \left|E_\mathrm{cam}(x,y;\vec{u})\right|^2 \, \dd x\,\dd y.
\end{align}
Thus, the lens-based optical front end, together with the EMCCD pixel readout, defines a deterministic mean input-to-readout map $\vec{u}\mapsto \vec{x}(\vec{u})$, while the experimentally observed outputs fluctuate around this mean because of measurement noise.

\subsection{\label{si:lens_noise}Noise in lens-based setup, photon budget estimation, and thresholding as a preprocessing method}

\subsubsection{EMCCD measurement model}

Scientific cameras produce quantitative images by counting photons that interact with the detector. Each sensor consists of millions of pixels that record both photon location and photon number, yielding a digital bitmap. The pixel value $X_\mathrm{ph}$ produced by the sensor in response to incident photons results from a sequence of stochastic processes~\cite{basden2003photona, lantz2008multiimaging, hirsch2013stochastic, toninelli2020quantum}. The number of incident photons over a fixed exposure follows a Poisson distribution, and each photon is independently converted into a photoelectron with probability $\eta$ (the quantum efficiency). In an EMCCD, photoelectrons are then amplified along a multiplication register via impact ionization; in the large-gain limit, the number of output electrons per input photoelectron is approximately exponentially distributed with mean $g_\mathrm{EM}$. Finally, the output electron count is mapped linearly by an analog-to-digital converter (ADC) to an integer pixel value (in units of the gray level, gl), with proportionality $g_\mathrm{ADC}$. The overall gain $g \equiv g_\mathrm{EM}\,g_\mathrm{ADC}$ relates the mean photoelectron number to the mean photon-induced pixel value.

Considering all the above processes, the distribution of $X_\mathrm{ph}$ can be approximated well by the following continuous probability distribution function (PDF) in the large gain limit~\cite{hirsch2013stochastic}:
\begin{equation}
    f(X_\mathrm{ph}|\lambda, \eta, g)=\sqrt{\frac{\eta\lambda}{g X_\mathrm{ph}}}I_1\qty(2\sqrt{\frac{\eta\lambda X_\mathrm{ph}}{g}})e^{-X_\mathrm{ph}/g-\eta\lambda},
\end{equation}
where $\lambda$ is the mean of the incident photon number during the exposure, $\eta$ is the quantum efficiency, $g$ is the gain, defined as ratio between mean gray level and mean photoelectron number, and $I_1$ is the modified Bessel function of the first kind. The mean and the variance of this distribution are:
\begin{equation}
    \mean[X_\mathrm{ph}]=\eta\lambda g,\quad\Var[X_\mathrm{ph}]=2\eta\lambda g^2.
    \label{sieq:var_mean}
\end{equation}

In the absence of incident light, the gray levels are still nonzero due to the presence of dark noise and offset. Nonzero gray levels, or equivalently, nonzero electrons, mainly come from three sources \cite{lantz2008multiimaging, toninelli2020quantum}: zero-mean Gaussian readout noise; clock-induced charge (CIC), which after EM gain produces an approximately exponential tail; and thermal dark current, which is strongly suppressed by detector cooling. An electronic offset is added so that measured pixel values remain non-negative even in the presence of zero-mean readout fluctuations.

\SuppEqref{sieq:var_mean} provides the moment relation used for the trace-based gain calibration below. The dark-shot histogram fit described later is used to define an optional thresholding preprocessing step and is not used to estimate the gain.

\subsubsection{Trace-based mean-variance EMCCD gain calibration and photon budget estimation}

For each lens-based dataset, the EMCCD gain was calibrated using the mean-variance relation \suppEqref{sieq:var_mean} based on repeated shots. The calibration unit was a pixel trace: for a fixed input image and a fixed camera pixel, we used the sequence of raw pixel values measured over the $S_\mathrm{max}$ repeated shots. This keeps the incident intensity approximately fixed while enabling us to measure the shot-to-shot fluctuations of the EMCCD readout.

For each experiment, we acquired $S_\mathrm{max}$ shots of dark frames with the same exposure time and EM-gain settings as the corresponding bright shots, which is 100 dark shots for each power level of MNIST data and 300 dark shots for MPEG-7 data. These dark shots were used to estimate the empirical dark mean $\bar{X}_\mathrm{dark}$, which was subtracted from the calibration-bin means. 

Pixel traces were grouped into bins according to their shot-averaged pixel values. For each bin, we randomly sampled up to 100 traces. The bin mean $m_j$ was computed as the average shot-averaged pixel value of the selected traces, while the bin variance $v_j$ was computed as the average within-trace shot variance. Prior to computing bin statistics, a trace-level quality control step removed traces with anomalously high within-trace variance or near-saturation pixel values; these accounted for less than 0.1\% of all traces, and arose from rare camera readout artifacts, hot pixels, or mechanical and electronic glitches during acquisition. Uncertainties on the bin statistics were estimated by hierarchical bootstrap resampling—of traces, and of shots within each selected trace—using 2{,}000 replicates per dataset.

From the bin statistics, the gain was then obtained from a weighted linear fit
\begin{equation}
    \hat{v}_j=a(\hat{m}_j-\bar{X}_\mathrm{dark})+b,
\end{equation}
using the bootstrap standard error of $\hat{v}_j$ as the $y$ uncertainty. From the EMCCD moment relation, $\Var[X_\mathrm{ph}]=2g\mean[X_\mathrm{ph}]$, the gain was obtained as $g=a/2$. The resulting gains were $g=54.57\pm0.36$ for MNIST low power, $g=58.80\pm0.33$ for MNIST high power, and $g=167.04\pm0.35$ for MPEG-7. 

Photon budgets were estimated from raw, non-thresholded, shot-averaged data. After subtracting the empirical dark-shot mean $\bar{X}_\mathrm{dark}$, the detected photon budget per input and per shot was estimated by summing the background-subtracted averaged frame and dividing by the calibrated gain $g$:
\begin{equation}
    N_\mathrm{ph} = \frac{1}{N_\mathrm{input} g} \sum_{n=1}^{N_\mathrm{input}} \sum_{k=1}^{K} \big(\bar{X}_k(\vec{u}_n) - \bar{X}_\mathrm{dark}\big),
\end{equation}
giving detected photon budgets of $33.1\pm 0.2$, $696\pm 4$, and $798\pm 2$ per shot for MNIST low power, MNIST high power, and MPEG-7, respectively, where the standard deviations across input images were $8$, $29$, and $46$. Using $\eta\approx65\%$ at $\lambda=\SI{355}{nm}$~\cite{ProEM_HS_512BX3}, the corresponding incident photon budgets are $51$, $1071$, and $1227$ photons per shot, respectively.

\subsubsection{Dark-shot histogram fitting and thresholding as preprocessing}

\begin{figure}[!p]
    \centering
    \includegraphics[scale=\scale]{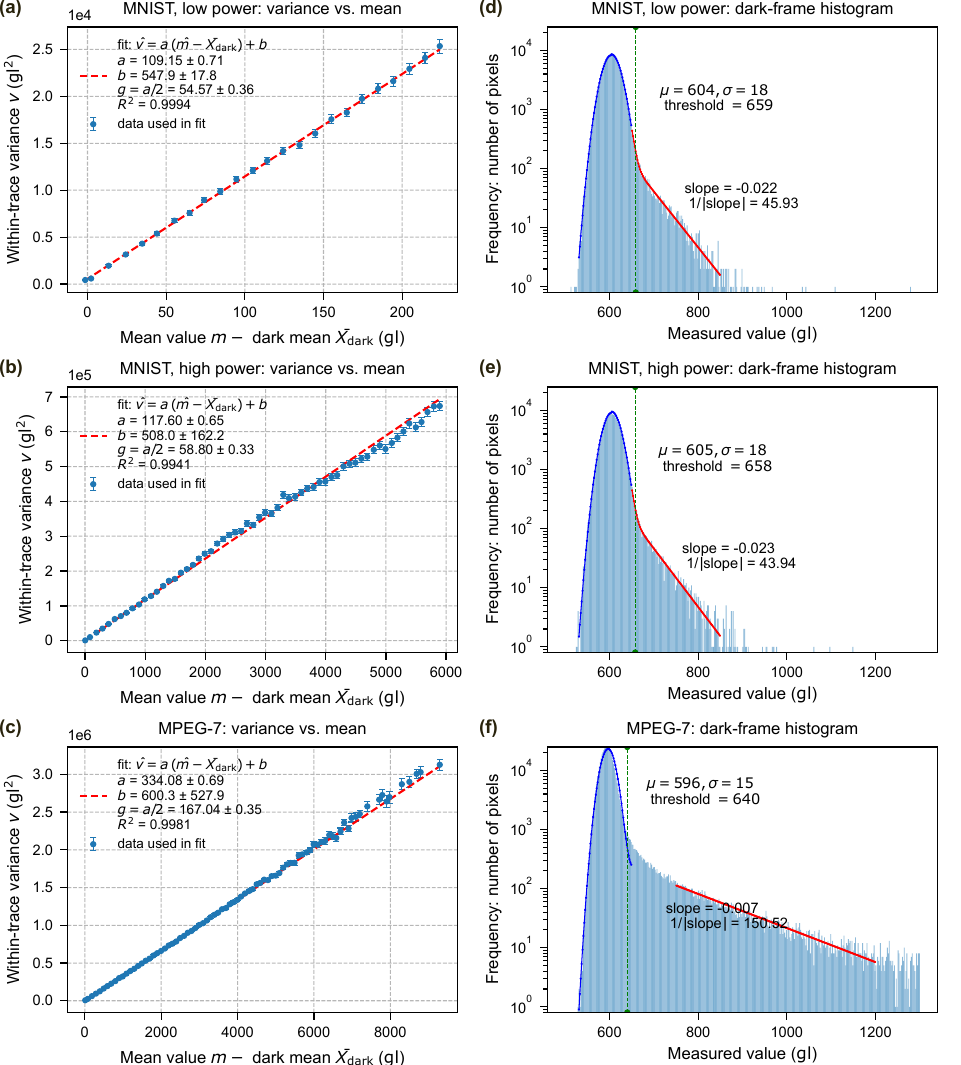}
    \caption{
    \figtitle{Calibration of EMCCD camera.}
    \figpanel*{a}{c} Trace-based mean-variance calibration for MNIST low-power data, MNIST high-power data, and MPEG-7 data, respectively. Repeated measurements of the same input and camera pixel were grouped by their shot-averaged pixel values. For each bin, the plotted variance is the average within-trace shot variance, and error bars are hierarchical-bootstrap standard errors. Red dashed lines show weighted linear fits $\hat v_j=a(\hat m_j-\bar{X}_\mathrm{dark})+b$, where $\bar{X}_\mathrm{dark}$ is the empirical dark-shot mean. The gain is $g=a/2$. 
    \figpanel*{d}{f} Dark-shot histograms measured with matched camera settings. Blue curves show Gaussian fits to the readout-noise peak, red curves show exponential tail fits, and green dashed lines indicate the optional threshold $\mu+3\sigma$. The fitted $\mu$ and $\sigma$ were used only for thresholding preprocessing, not for the gain calibration in \figpanel*{a}{c}.
    }
    \label{sifig:emccd_power}
\end{figure}

The same dark shots were also used to define the thresholding preprocessing applied in the classification experiments. The histogram of dark-shot pixel values was fitted phenomenologically by a Gaussian readout-noise peak plus an exponential tail,
\begin{equation}
    h(x)
    =
    A\exp\left[
        -\frac{(x-\mu)^2}{2\sigma^2}
    \right]
    +
    \exp(a_\mathrm{tail}x+b_\mathrm{tail})\mathbf{1}_{x\ge \mu},
\end{equation}
where $x$ is the measured pixel value, $\mu$ is the fitted Gaussian peak center, $\sigma$ is the fitted readout-noise width, and $a_\mathrm{tail}$ and $b_\mathrm{tail}$ describe the exponential tail associated primarily with CIC and dark events after EM amplification. The fitted results are shown in \suppfigpanelref*{sifig:emccd_power}{d}{f}. We emphasize that the fitted Gaussian dark-peak center $\mu$ is distinct from $\bar{X}_\mathrm{dark}$, the mean of the dark-shot data.

From dark noise distribution, the pixel values within $\mu\pm3\sigma$ are dominated by dark noise rather than photon-induced signal. In the large-gain regime, retaining such values therefore contributes more noise than information. We accordingly apply an optional thresholding step that subtracts the dark peak mean $\mu$ and zeros out any residual value below $3\sigma$.
This preprocessing step was used only in the thresholded-data classification experiments reported in \suppnoteref{si:thresholding_compare}; it was applied neither in the trace-based gain calibration and photon-budget estimates reported above, nor in the main-text experiments reported in Section \ref{sec:result}.

\FloatBarrier

\section{\label{si:mnist_lens}Supplementary results of MNIST classification with lens-based front end}
\subsection{\label{si:eigentask_vis}Features, eigentasks and their visualization}
\begin{figure}[!htbp]
    \centering
    \includegraphics[scale=\scale]{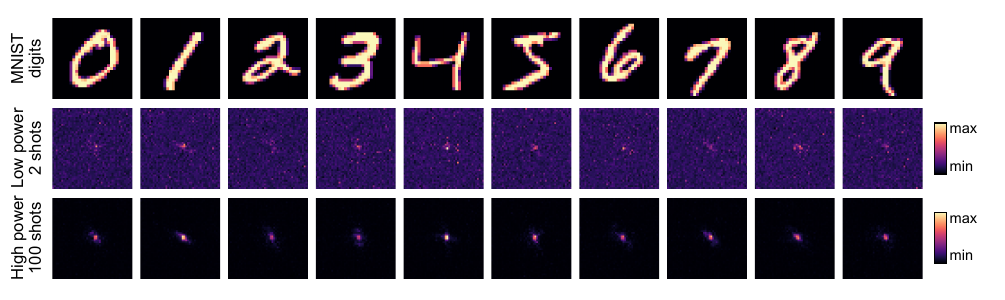}
    \caption{
    \figtitle{MNIST digits and their measured features.} MNIST digits (first row) and averaged frames used as features from low-power data averaged over 2 shots (second row) and high-power data averaged over 100 shots (third row). The low-power data is significantly noisier than the high-power data.}
    \label{sifig:mnist_features}
\end{figure}

The original MNIST digits and features are shown in \suppfigref{sifig:mnist_features}. The second line is the feature obtained by averaging over 2 shots of low-power data. The third row shows features obtained by averaging more than 100 shots of high-power data. The second row is noisier than the third row because fewer shots are averaged, leaving larger EMCCD sampling fluctuations as described in \suppnoteref{si:lens_noise}.

\begin{figure}[!p]
    \centering
    \includegraphics[scale=\scale]{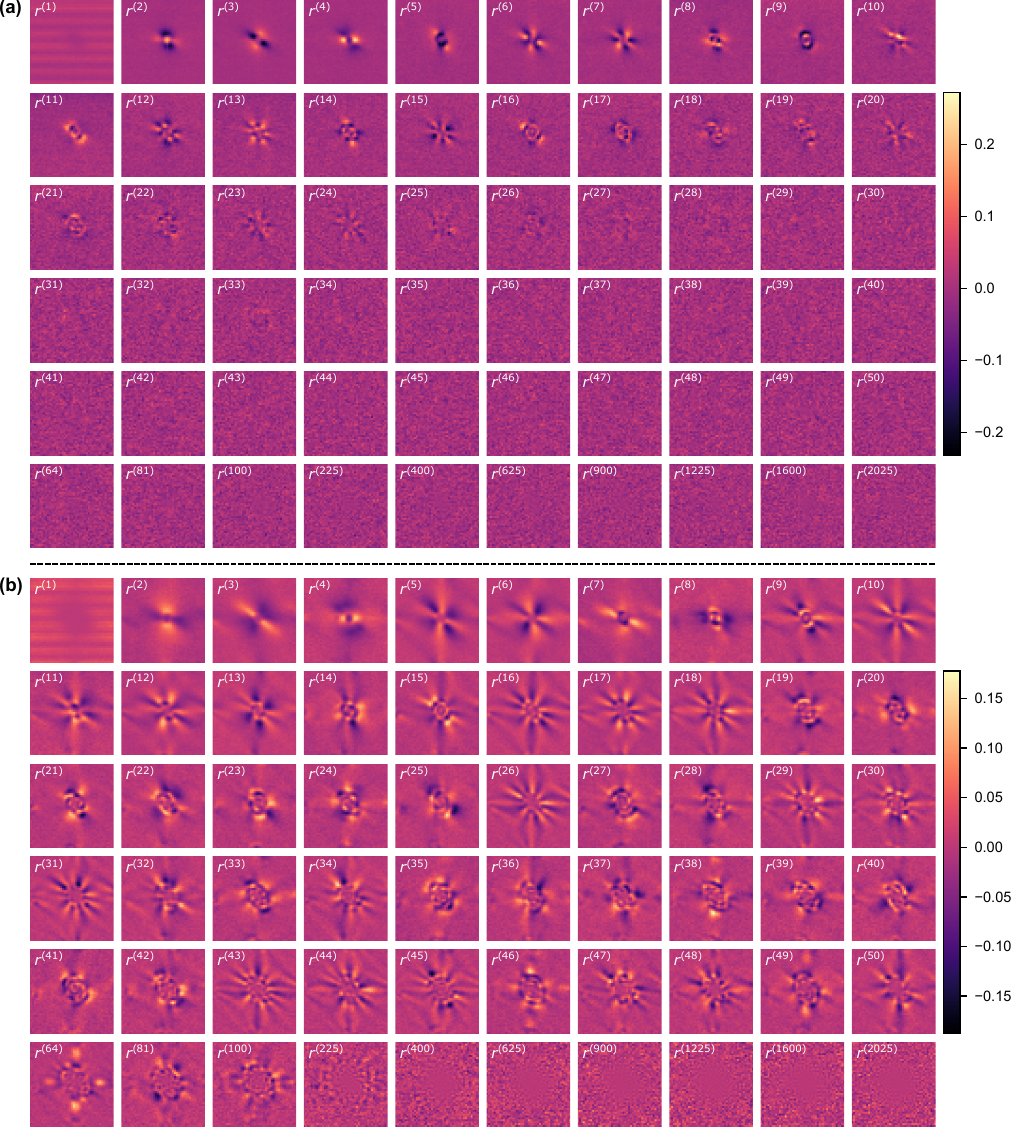}
    \caption{\figtitle{MNIST eigentask weights.}
    \figpanel{a}{b} Eigentask weights $\vec{r}^{(k)}~(k=1, 2 \dots, 49, 50, 64, 81, 100, 225, 400, 625, 900, 1225, 1600, 2025)$ estimated from 100 shots in low-power and high-power data, respectively.}
    \label{sifig:mnist_eigen_masks}
\end{figure}

The eigentask weights, $\vec{r}$ in \eqnref{eq:eigentaskProb}, are plotted in \suppfigref{sifig:mnist_eigen_masks}. \Suppfigpanelref{sifig:mnist_eigen_masks}{a}{b} are eigentasks for low-power and high-power datasets, respectively. For each dataset, eigentasks are estimated from 100 shots from the training set. The first 50 weights, $\vec{r}^{(k)}~(k=1, 2, \dots, 50)$, as well as $k=L^2~(L=8, 9, 10, 15, 20, 25, 30, 35, 40, 45)$, are plotted after being reshaped into 2D arrays. As the order $k$ increases, eigentask weights are getting more noisy, until signals are submerged in noises at some $k$. For low-power data, there are about 25-30 clearer eigentasks, while for high-power data there are about 150-200. The first set of weights, $\vec{r}^{(1)}$, is near-uniform, consistent with the approximately conserved total optical power in the phase-only, focal-plane readout described in \suppnoteref{si:lens_transform}.

\clearpage

\subsection{\label{si:mnist_high_power}Classification on high-power data}

\begin{figure}[!htbp]
    \centering
    \includegraphics[scale=\scale]{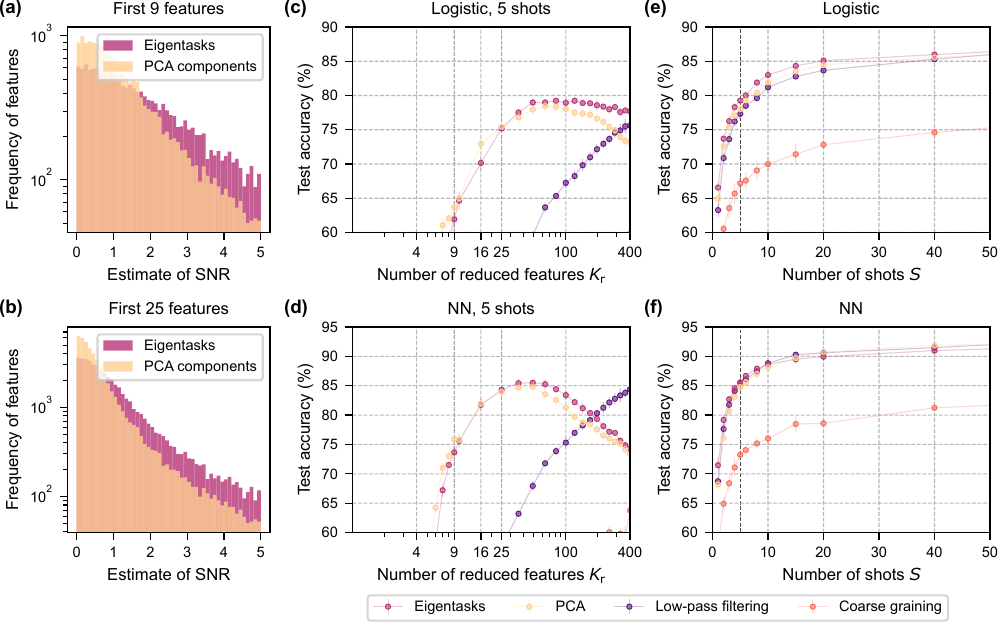}
    \caption{
    \figtitle{Evaluation of noise-mitigation techniques with MNIST classification for high-power data in a simple lens-based optical front end.} 
    \figpanel{a}{b} Histograms of an empirical per-feature SNR (estimated from 100-shot data) for the first 9 or 25 retained PCA features and the leading 9 or 25 non-normalization eigentask features.
    \figpanel{c}{d} Test accuracy versus reduced dimension $\Kr$ at $S=5$ shots for logistic regression and neural network, respectively, comparing the four transforms. The two vertical dashed lines indicate the positions of 9 and 25 features in panel \figpanel{a} and \figpanel{b}.
    \figpanel{e}{f} Test accuracy versus number of shots $S$ for logistic regression and neural network, respectively. The vertical dashed line indicates the position of $S=5$ shots plotted in \figpanel{c} and \figpanel{d}. 
Model selection is described in Methods. Accuracy panels are means over five independent training/validation/test splits; error bars denote one standard deviation. Histograms are shown for a representative split.
    }
    \label{sifig:mnist_high_power}
\end{figure}

The classification results on high-power data are shown in \suppfigref{sifig:mnist_high_power}. 

\Suppfigpanelref{sifig:mnist_high_power}{a}{b} shows the histograms of single-feature SNRs, estimated from 100-shot data, for the noise-mitigated features ${\vec{Y}}_\mathrm{EGT}$ and $\vec{Y}_\mathrm{PCA}$  for the first 9 and 25 features of each input digit. As in the low-power case, eigentasks exhibit a heavier high-SNR tail than PCA components in both cases, indicating eigentasks' ability to extract less noisy features. At the same time, the overall SNR distribution are shifted to higher values than in the low-power setting, which is consistent with the cleaner data obtained at high optical power, while the relative advantage of eigentasks over PCA remains visible.

In \suppfigpanelref{sifig:mnist_high_power}{c}{d}, we evaluate the test accuracy as a function of the number of reduced features $\Kr$ for 5-shot-averaged high-power data with \figpanel{c} using logistic regression and \figpanel{d} using a neural network. Similar to the low-power case, We again observe an overfitting trend characteristic of supervised learning~\cite{hughes1968mean, trunk1979problem, hua2005optimal}: increasing $\Kr$ beyond an intermediate range does not continue to improve performance, as high-order features provide more noise than information and don't help in the training. The maximum accuracy values of eigentask-based classification are achieved in a broader optimal range near $\Kr\approx 60\text{--}100$, which is consistent with \suppfigref{sifig:mnist_eigen_masks}, where high-power eigentask weights remain visually structured to similar and higher orders.

In \suppfigpanelref{sifig:mnist_high_power}{e}{f}, we present test accuracy versus shot count $S$ for \figpanel{e} logistic regression and \figpanel{f} neural network classifiers. As expected, all methods improve with increasing $S$ as sampling noise is reduced. Compared with the low-power results, the overall accuracy is higher and the gap between different noise-mitigation methods is smaller, reflecting the reduced impact of sampling noise at high power. Eigentasks still attain the best or near-best performance among the four methods, while more advantage is observed in the few-shot regime, confirming our observation that the main benefit of eigentask learning is to preserve strong performance at low sampling budgets without sacrificing accuracy in the many-shot limit.

\subsection{\label{si:thresholding_compare}Learning in the presence of thresholding preprocessing}

\begin{figure}[!htbp]
    \centering
    \includegraphics[scale=\scale]{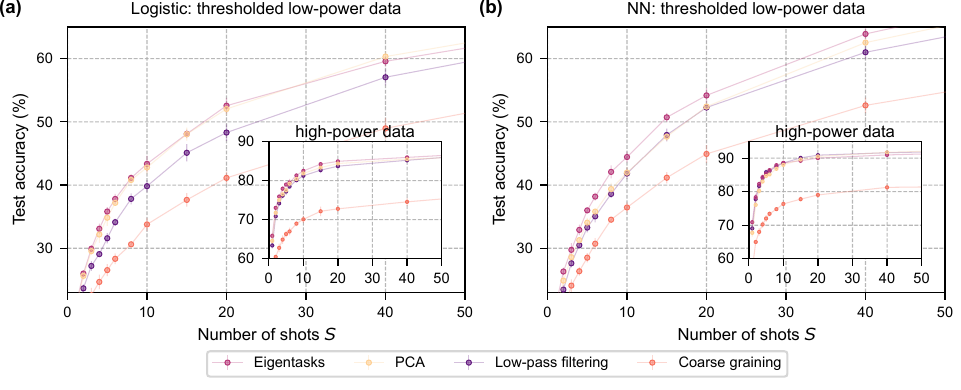}
\caption{\figtitle{Performance of noise-mitigation techniques for MNIST classification with thresholded data using a simple lens-based optical front end.} \figpanel{a}{b} Test accuracy versus number of shots $S$ for logistic regression and a neural-network classifier, respectively, on thresholded low-power data; insets show the corresponding results for thresholded high-power data. 
Model selection is described in Methods. Data are means over five independent training/validation/test splits; error bars denote one standard deviation.
    }
    \label{sifig:mnist_thresholded}
\end{figure}

We next examine classification accuracy as a function of the number of shots $S$ for thresholded data and compare the four noise-mitigation techniques across power levels; the results are shown in \suppfigref{sifig:mnist_thresholded}.

For the thresholded low-power data shown in the main panels of \suppfigref{sifig:mnist_thresholded}, eigentask learning remains the best or near-best method across the tested shot range. The advantage is clearest for the neural-network classifier and in the few-shot regime for logistic regression, while at larger $S$ the competing methods become closer, as expected when sampling noise is reduced. Qualitatively, this behavior is consistent with \figref{fig:mnist_main}: thresholding does not alter the overall ordering of methods or remove the benefit of eigentask-based noise-mitigation. Instead, the results indicate that the eigentask advantage is not tied to a particular choice of raw versus thresholding preprocessing, but reflects its ability to construct a measurement-adapted and noise-robust feature representation.

The same qualitative ordering is retained for thresholded high-power data, as shown in the insets of \suppfigref{sifig:mnist_thresholded}. In this high-SNR regime, the gap between methods narrows, consistent with \suppfigref{sifig:mnist_high_power}, but eigentask learning still remains best or near-best. Taken together, these results show that thresholding does not qualitatively change the conclusions of the MNIST lens-front-end study: eigentasks remain competitive across preprocessing choices and power levels, with the strongest gains appearing in the more noise-dominated regimes.

\FloatBarrier
\clearpage

\section{\label{si:spdnn_encoder}Details of SPDNN encoder}
\subsection{\label{si:spdnn_setup}Details of the SPDNN experimental setup}

\begin{figure}[!htbp]
    \centering
    \includegraphics[width=\textwidth]{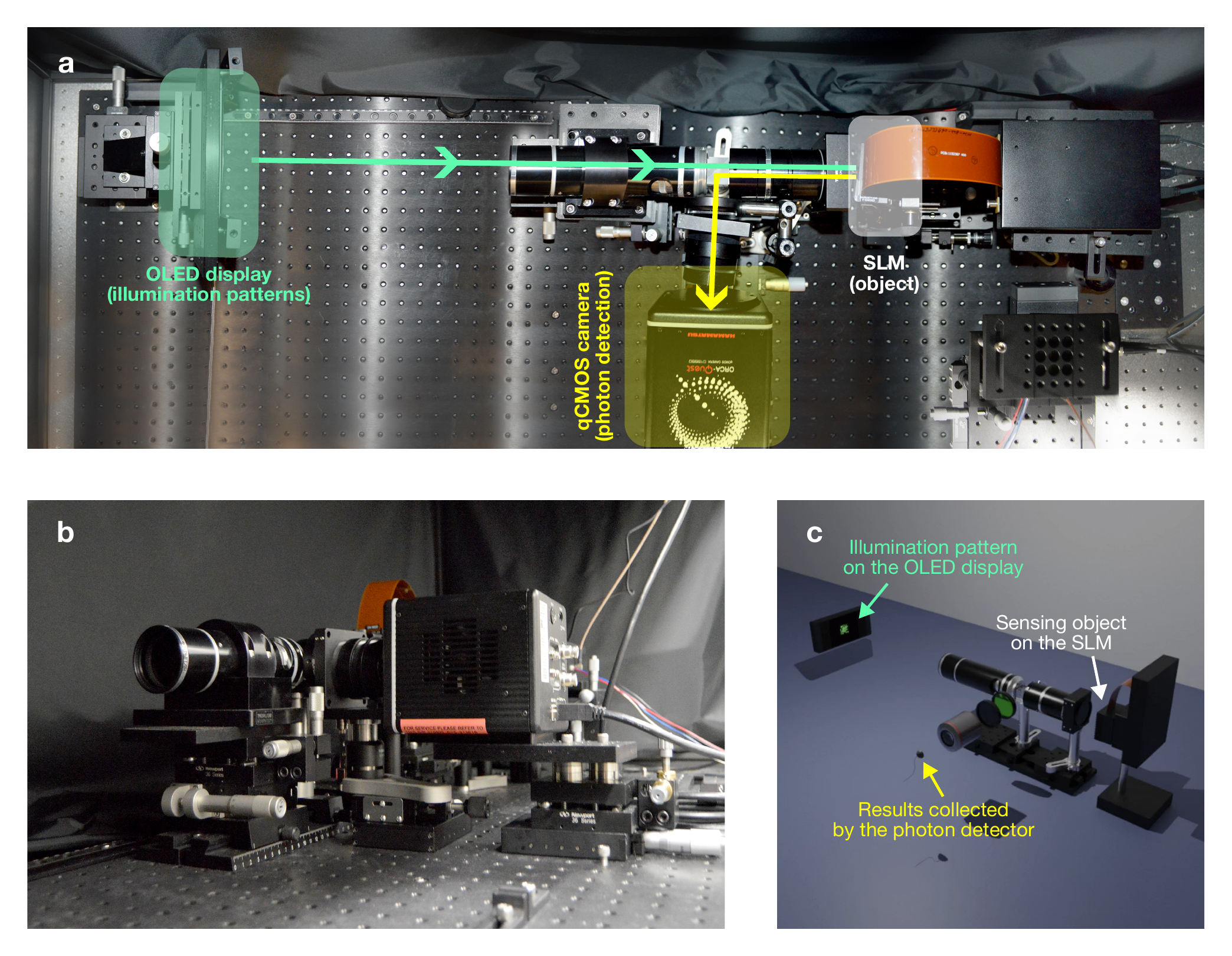}
    \caption{\figtitle{SPDNN experimental setup.} \figpanel{a} Top-down photograph of the optical system with the three main components highlighted: OLED display for programmable illumination patterns (green), SLM for object emulation (white), and qCMOS camera for single-photon detection (yellow). Arrows indicate the light path from illumination source through the object plane to the detector. \figpanel{b} Side-view photograph of the setup. \figpanel{c} Schematic rendering illustrating the sensing pipeline: illumination patterns displayed on the OLED probe the digit on the SLM, and the transmitted light is measured by the photon detector.}
    \label{sifig:SPDNN_setup}
\end{figure}

\Suppfigref{sifig:SPDNN_setup} shows photographs and a schematic rendering of the apparatus. The system consists of three functional stages: (1)~a programmable organic light-emitting diode (OLED) display encoding the illumination patterns, (2) an SLM serving as the object plane where illumination interacts with the sample, and (3)~a quantitative scientific CMOS (qCMOS) camera performing single-photon detection. As illustrated by the light path in \suppfigpanelref{sifig:SPDNN_setup}{a}, light propagates from the OLED through a zoom lens system to the SLM, then through polarization optics and spectral filtering before reaching the camera.

\subsection{\label{si:spdnn_setup_noise}SPDNN encoder and its sampling noise}

We next summarize the single-photon-detection neural network (SPDNN) encoder used in the analysis of \figref{fig:spdnn_main}. The underlying setup and its experimental validation are described in Ref.~\cite{ma2025quantumlimited}; here we retain only the ingredients needed to define the effective input-to-readout map and the associated sampling noise.

The SPDNN considered here is the incoherent MNIST model of Ref.~\cite{ma2025quantumlimited}. An input digit is represented by a non-negative vector $\vec u \in [0,1]^{784}$. We use the first hidden layer of width $K$ as the optical encoder. Its pre-activation vector is
\begin{align}
    \vec z(\vec u) = \mat{M} \vec u ,
\end{align}
with $\mat{M}$ being the matrix implemented by optical matrix-vector multiplier (MVM) and $M_{ij}\ge 0$ in the incoherent implementation. As in Ref.~\cite{ma2025quantumlimited}, $z_k$ is interpreted in units of mean detected photons per shot at the $k$-th single-photon detector (SPD), up to an overall calibration factor that may be absorbed into $\mat{M}$. Each SPD produces a binary output in one shot: either no click or click. For Poissonian illumination with mean photon number $\lambda_k(\vec u)=z_k(\vec u)$, the click probability is
\begin{align}
    p_k(\vec u)=P_{\mathrm{SPD}}(\lambda_k)=1-e^{-\lambda_k(\vec u)}
    =1-e^{-z_k(\vec u)} .
\end{align}
Accordingly, the single-shot encoder output at neuron $k$ is a Bernoulli random variable,
\begin{align}
    X_k^{(s)}(\vec u)\sim \mathrm{Bernoulli}\!\left(p_k(\vec u)\right)\quad\Leftrightarrow\quad X_k^{(s)}(\vec u)=
    \begin{cases}
        1, & \mathrm{with probability } p_k(\vec u),\\
        0, & \mathrm{with probability } 1-p_k(\vec u).
    \end{cases}
\end{align}
Its mean and variance are therefore
\begin{align}
    \mean[X_k^{(1)}(\vec u)] = p_k(\vec u),\quad    \Var[X_k^{(1)}(\vec u)] = p_k(\vec u)\bigl(1-p_k(\vec u)\bigr).
\end{align}
For an $S$-shot averaged readout $\bar X_{k,S}(\vec u)=\frac{1}{S}\sum_{s=1}^{S} X_k^{(s)}(\vec u)$ as defined previously, we have:
\begin{align}
    \mean[\bar X_{k,S}(\vec u)] = p_k(\vec u),\quad\Var[\bar X_{k,S}(\vec u)]= \frac{p_k(\vec u)\bigl(1-p_k(\vec u)\bigr)}{S}.
\end{align}
Thus, the SPDNN features used in our eigentask analysis are the hidden-neuron click outcomes, or their shot-averaged values, rather than raw camera-pixel intensities.

In this work, the eigentask and PCA bases for the SPDNN encoder are learned from simulated SPDNN outputs generated with the validated model of Ref.~\cite{ma2025quantumlimited}, and then evaluated on both simulated data and the 100-digit experimental dataset reported there.

\subsection{\label{si:spdnn_transform}Optical encoding map implemented by the SPDNNs}

Unlike the lens-based case, where the optical front end induces a fixed optical encoding map, the SPDNN encoder is trained and task-specific. For the MNIST models analyzed here, we treat only the first hidden layer of the SPDNN as the optical encoder. The relevant input-to-feature map is therefore
\begin{align}
    \vec u \mapsto \vec z(\vec u)=\mat{M}\vec u \mapsto \vec x(\vec u),
\end{align}
where $\vec z(\vec u)\in\mathbb{R}_{\ge 0}^{N}$ denotes the pre-activation vector and $\vec x(\vec u)$ denotes the mean encoder output. For the incoherent SPDNN, we have (see \suppnoteref{si:spdnn_setup_noise}):
\begin{equation}
    \vec{x}(\vec{u})=P_\mathrm{SPD}(\mat{M}\vec{u})=1-e^{-\mat{M}\vec{u}}.
\end{equation}

For clarity, we emphasize that Ref.~\cite{ma2025quantumlimited} completes the original SPDNN classifier with a second, high-SNR output layer. In the present work, however, that output layer is not part of the encoder definition. Instead, the hidden-layer SPD readouts themselves are treated as the measured features supplied to the downstream noise-mitigation and classification pipeline.
\section{\label{si:spdnn_results}Supplementary results of MNIST classification with SPDNN encoder}
\subsection{\label{si:spdnn_sim}Classification on larger test set by simulation}

\begin{figure}[!t]
    \centering
    \includegraphics[scale=\scale]{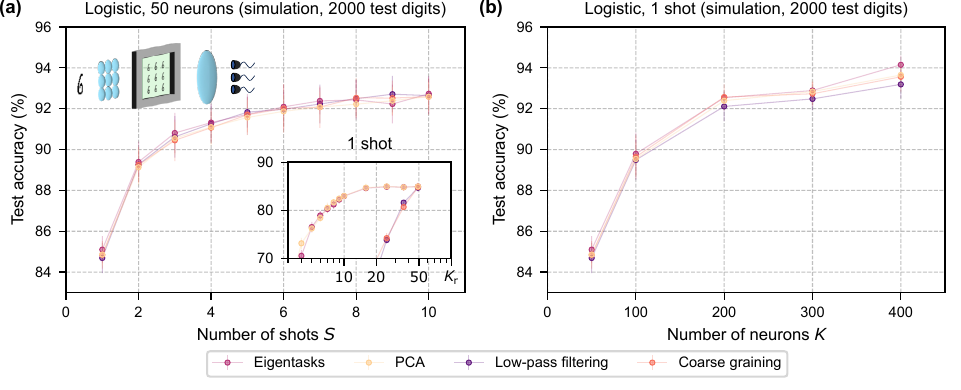}
    \caption{
    \figtitle{Performance of noise-mitigation techniques in MNIST classification using SPDNN simulation data with logistic regression.}
    \figpanel{a} Test accuracy on 2{,}000 test digits in simulation versus number of shots $S$ for a 50-neuron SPDNN. The inset shows accuracy versus the reduced feature dimension $\Kr$ at $S=1$.
    \figpanel{b} Test accuracy on the same 2{,}000 digits versus neuron count $K$ for simulated data at $S=1$ shot. Model selection is described in Methods. Data are means over five independent simulated training/validation/test subsets; error bars denote one standard deviation.
    }
    \label{sifig:spdnn_sim}
\end{figure}

In the main text, we presented the classification performance of various noise-mitigation methods on only 100 test digits, limited by the experimental data provided in Ref.~\cite{ma2025quantumlimited}. To verify that the advantage of eigentask learning is not confined to this small test set, we further evaluated the same SPDNN pipeline on a larger simulated test set of 2{,}000 digits with logistic classifiers, with the results shown in \suppfigref{sifig:spdnn_sim}.

\Suppfigpanelref{sifig:spdnn_sim}{a} shows the test accuracy as a function of the shot count $S$ for a 50-neuron SPDNN on 2{,}000 simulated test digits. In contrast to \figpanelref{fig:spdnn_main}{a}, where the performance on 100 experimental test digits quickly approaches $95\%$, the accuracy here remains at a lower level with fewer fluctuations. Eigentasks still consistently perform better than the other methods in few-shot regime. As in all previous results, the accuracy increases with the shot count $S$, reflecting the reduced stochastic noise at larger shot counts. The inset shows the test accuracy as a function of the number of reduced features $\Kr$ for single-shot testing, exhibiting a similar overfitting behavior as in the previous figures.

\Suppfigpanelref{sifig:spdnn_sim}{b} presents the test accuracy as a function of the number of neurons $K$ for single-shot data. Consistent with \figpanelref{fig:spdnn_main}{b} with a smaller test set size, the accuracy improves as the neuron count increases, indicating that larger SPDNNs provide more useful information via more features for the classifiers. Eigentasks remain the best or near-best method across the tested neuron counts.

Together with \figref{fig:spdnn_main}, these results show that the performance gain from eigentask learning is not specific to a small set of 100 digits, but remains on a substantially larger set as well.

\subsection{\label{si:spdnn_nn}Classification with neural networks}

Besides logistic regression, we also performed classification on the SPDNN data using a neural-network back-end classifier, for both the experimental data from Ref.~\cite{ma2025quantumlimited} and the larger simulated test set reported in \suppnoteref{si:spdnn_sim}.

\begin{figure}[!t]
    \centering
    \includegraphics[scale=\scale]{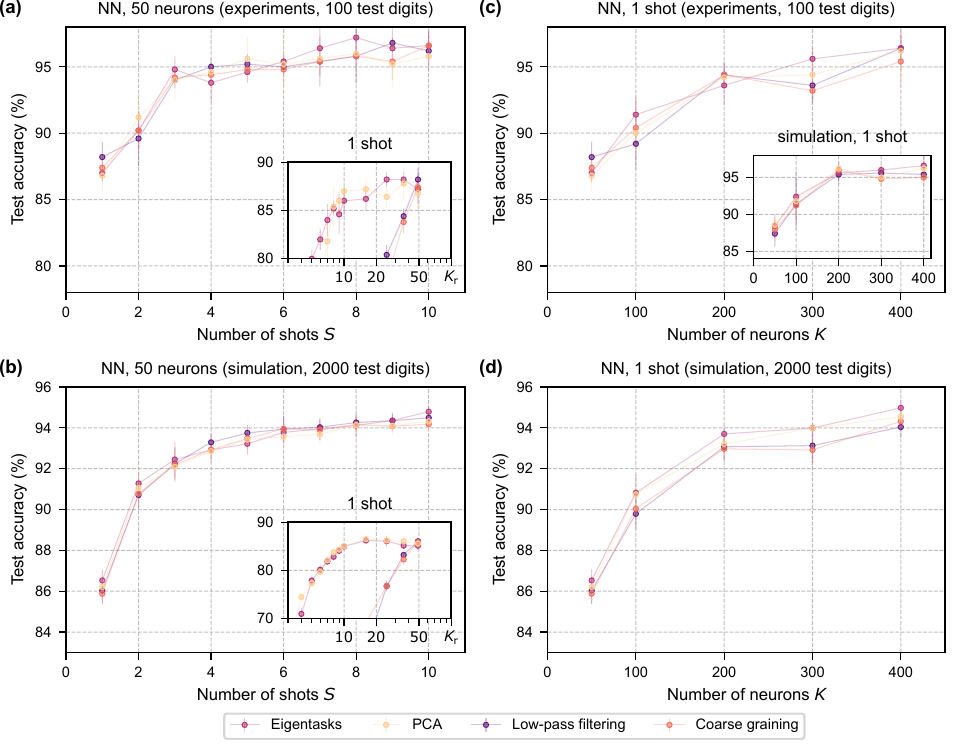}
    \caption{\figtitle{Performance of noise-mitigation techniques in MNIST classification using SPDNN data and neural networks with a hidden layer of 400 neurons.}
    \figpanel{a}{b} Test accuracy on 100 test digits in experiments and 2{,}000 test digits in simulation, respectively, versus number of shots $S$ for a 50-neuron SPDNN. The inset shows accuracy versus the reduced feature dimension $\Kr$ at $S=1$.
    \figpanel{c}{d} Test accuracy on 100 test digits in experiments and 2{,}000 test digits, respectively, versus neuron count $K$ at $S=1$ shot. Model selection is described in Methods. Data are means over five independent seeds. For the 100-digit experimental tests, the test set is fixed and only the simulated training/validation subsets vary across seeds; for the 2{,}000-digit simulation tests, the simulated training/validation/test subsets are independently sampled for each seed. Error bars denote one standard deviation.}
\label{sifig:spdnn_nn}
\end{figure}

\Suppfigpanelref{sifig:spdnn_nn}{a}{b} show the neural-network test accuracy as a function of the shot count $S$ for a 50-neuron SPDNN, with \figpanel{a} 100 test digits and \figpanel{b} 2{,}000 test digits. For the 100-digit experimental test set in \suppfigpanelref{sifig:spdnn_nn}{a}, similar to \figpanelref{fig:spdnn_main}{a}, the accuracy saturates rapidly. However, the neural-network classifier narrows the gap between eigentasks and other methods even in few-shot regime. Eigentask learning still performs the best in the few-shot regime of $S=1\text{--}3$ and achieves best or near-best accuracy for larger shot count.

\Suppfigpanelref{sifig:spdnn_nn}{c}{d} show the test accuracy as a function of neuron count $K$ for single-shot data, with \figpanel{c} 100 test digits and \figpanel{d} 2{,}000 test digits. For the 100-digit experimental test set in \suppfigpanelref{sifig:spdnn_nn}{c}, all noise-mitigation methods achieve around $90\%$ or higher accuracy with only $1$ shot. However, eigentasks are not uniformly superior at every data point: low-pass filtering, PCA, or coarse graining can match or slightly exceed eigentasks at some $K$, though the differences are generally within the error bars. For the larger and more statistically robust simulated test set with 2{,}000 digits in \suppfigpanelref{sifig:spdnn_nn}{d}, the accuracy improves with increasing $K$ and saturates to $97.2\%$. Eigentasks attain the best or near-best results across all neuron counts, but the lead is marginal.

In summary, \suppfigref{sifig:spdnn_nn} shows that while eigentask learning remains a robust information extraction method for noisy SPDNN data, it is less advantageous in the regime where the optical encoder is already task-optimized and the classifier is sufficiently expressive. The advantage of eigentasks on noisy data over other methods can be partially compensated for by additional optimization elsewhere in the system when ample computational resources are available. In this regime, eigentask learning should be seen as a competitive and reliable method rather than as a universally dominant one.

\putbib[bibfile]
\end{bibunit}

\end{document}